\documentclass[fleqn,usenatbib]{mnras}

\usepackage{newtxtext,newtxmath}

\usepackage[T1]{fontenc}
\usepackage{ae,aecompl}

\usepackage{graphicx}	
\usepackage{amsmath}	
\usepackage{amssymb}	
\usepackage{hyperref}

\defcitealias{Mertens2018}{M18}
\defcitealias{Mertens2020}{M20}
\defcitealias{Gosh2020}{G20}

\def\para{\parallel}
\def\d{\boldsymbol{d}}

\def\e{\boldsymbol{e}}
\def\E{\boldsymbol{E}}
\def\f{\boldsymbol{f}}
\def\F{\boldsymbol{F}}
\def\H{\boldsymbol{H}}
\def\r{\boldsymbol{r}}
\def\R{\boldsymbol{R}}
\def\n{\boldsymbol{n}}
\def\N{\boldsymbol{N}}

\def\T{\boldsymbol{T}}

\def\g{\boldsymbol{g}}
\def\q{\boldsymbol{q}}
\def\c{\boldsymbol{c}}
\def\p{\boldsymbol{p}}

\def\M{\boldsymbol{M}}
\def\C{\boldsymbol{C}}
\def\W{\boldsymbol{W}}
\def\V{\boldsymbol{V}}
\def\Cfg{\boldsymbol{C}_{\rm fg}}
\def\Cto{\boldsymbol{C}_{\rm 21}}
\def\Cn{\boldsymbol{C}_{\rm n}}

\def\dip{\boldsymbol{d}_{\rm ip}}
\def\I{\boldsymbol{I}}
\def\Cov{{\rm Cov}}
\def\Exp{{\rm E}}
\def\bSigma{\boldsymbol{\Sigma}}
\def\bnu{\boldsymbol{\nu}}
\def\btheta{\boldsymbol{\theta}}
\def\K{\boldsymbol{K}}
\def\bk{\boldsymbol{k}}
\def\Kto{K_{\rm 21}}
\def\Kfg{K_{\rm fg}}
\def\Kn{K_{\rm n}}
\def\Kip{K_{\rm ip}}
\def\Km{K_{\rm Matern}}
\def\sigmafg{\sigma_{\rm fg}}
\def\ellfg{\ell_{\rm fg}}
\def\sigman{\sigma_{\rm n}}
\def\sigmato{\sigma_{\rm 21}}
\def\ellto{\ell_{\rm 21}}
\def\A{\boldsymbol{A}}
\def\B{\boldsymbol{B}}
\def\U{\boldsymbol{U}}
\def\S{\boldsymbol{S}}
\def\V{\boldsymbol{V}}

\def\tr{{\rm tr}}

\title[Gaussian Process Foreground Subtraction]
{Gaussian Process Foreground Subtraction and Power Spectrum Estimation for 21\,cm Cosmology}

\author[N. S. Kern \& A. Liu]{
Nicholas S. Kern$^{1}$\thanks{E-mail: nkern@mit.edu},
Adrian Liu$^{2}$
\\
$^{1}$Department of Physics and Kavli Institute for Astrophysics and Space Research, Massachusetts Institute of Technology, Cambridge, MA, USA \\
$^{2}$Department of Physics and McGill Space Institute, McGill University, 3600 University Street, Montreal, QC H3A 2T8, Canada
}

\date{Accepted XXX. Received YYY; in original form ZZZ}

\pubyear{2020}

\begin{document}
\label{firstpage}
\pagerange{\pageref{firstpage}--\pageref{lastpage}}
\maketitle

\begin{abstract}
One of the primary challenges in enabling the scientific potential of 21\,cm intensity mapping at the Epoch of Reionization (EoR) is the separation of astrophysical foreground contamination.
Recent works have claimed that Gaussian process regression (GPR) can robustly perform this separation, particularly at low Fourier $k$ wavenumbers where the EoR signal reaches its peak signal-to-noise ratio.
We revisit this topic by casting GPR foreground subtraction (GPR-FS) into the quadratic estimator formalism, thereby putting its statistical properties on stronger theoretical footing.
We find that GPR-FS can distort the window functions at these low $k$ modes, which, without proper decorrelation, make it difficult to probe the EoR power spectrum.
Incidentally, we also show that GPR-FS is in fact closely related to the widely studied optimal quadratic estimator.
As a case study, we look at recent power spectrum upper limits from the Low Frequency Array (LOFAR) that utilized GPR-FS.
We pay close attention to their normalization scheme, showing that it is particularly sensitive to signal loss when the EoR covariance is misestimated.
This has possible ramifications for recent astrophysical interpretations of the LOFAR limits, because many of the EoR models ruled out do not fall within the bounds of the covariance models explored by LOFAR.
Being more robust to this bias, we conclude that the quadratic estimator is a more natural framework for implementing GPR-FS and computing the 21\,cm power spectrum.
\end{abstract}




\section{Introduction}
\label{sec:intro}

The next generation of low frequency radio surveys promise to provide new insights into two key epochs in our universe's history:
the Cosmic Dawn (CD), which marks the formation of the first stars in the universe, and the subsequent Epoch of Reionization (EoR), which occured when the first stellar populations injected enough ultraviolet radiation into the surrounding intergalactic medium (IGM) to ionize its neutral hydrogen content \citep{Furlanetto2006c, Morales2010, Pritchard2012, Loeb2013}. 
Stellar feedback from these epochs had a profound impact on the state of the surrounding gas, and are therefore critically important to our broader understanding of galaxy and structure formation over cosmic time.
These radio experiments will use the redshifted 21\,cm line from neutral hydrogen in the IGM to systematically probe its density, ionization, and temperature state.
Due to cosmological redshift, we are able to relate the observed and rest-frame frequency of the 21\,cm line to deduce the redshift of its emission, thereby making it a powerful three-dimensional probe of IGM structure.
For a recent review of the state of 21\,cm cosmology, see \citet{Liu2020}.

Single-element radio telecsopes are aiming to measure the sky monopole component of the 21\,cm signal and its redshift evolution, which can constrain the timing and duration of the EoR and tell us when the first stars formed \citep[e.g.][]{Bowman2010, Singh2018}.
Recently the Experiment to Detect the Global EoR Signature (EDGES) experiment reported the first tentative detection of the 21\,cm absorption trough in the monopole component, or Global Signal, at Cosmic Dawn \citep{Bowman2018}.
The measured absorption signal was significantly deeper than expected, and may be a consequence of residual systematics \citep[e.g.][]{Bradley2019, Hills2018, Singh2019, Sims2020}; however, if proven to be truly cosmological in nature, the EDGES detection may be an indicator of a new paradigm of Cosmic Dawn astrophysics and fundamental physics \citep[e.g.][]{Barkana2018, Feng2018, Ewall-Wice2018, Munoz2018, Fialkov2019}.

The 21\,cm spatial fluctuations also contain a large amount of astrophysical information, which is encapsulated by its power spectrum.
The 21\,cm power spectrum is forecasted to be a powerful probe of IGM and galaxy formation astrophysics \citep{Pober2013b, Ewall-Wice2016a, Greig2016, Greig2017a}, and can be used in tandem with existing cosmological probes to more tightly constrain $\Lambda$CDM cosmology \citep{McQuinn2006, Mao2008, Liu2016a, Kern2017}. 
Over the past decade, interferometric experiments have placed increasingly stringent upper limits on the EoR 21\,cm power spectrum.
This includes the Giant Metrewave Radio Telescope \citep[GMRT;][]{Pen2009, Paciga2013}, the Donald C. Backer Precion Array for Probing the Epoch of Reionization \citep[PAPER;][]{Parsons2010, Jacobs2015, Kolopanis2019}, the Low Frequency Array \citep[LOFAR;][]{VanHaarlem2013, Mertens2020}, the Murchison Widefield Array \citep[MWA;][]{Bowman2013, Dillon2014, Ewall-Wice2016b, Beardsley2016, Barry2019a, Li2019, Trott2020}, and the Owens Valley Long Wavelength Array \citep[LWA;][]{Eastwood2019}.
Going forward, experiments like the Hydrogen Epoch of Reionization Array \citep[HERA;][]{DeBoer2017} and the Square Kilometer Array \citep[SKA;][]{Koopmans2015} aim to make high significance detections of the 21\,cm power spectrum, assuming systematics can be controlled.
HERA is particularly exciting in the near term as it is currently under construction, with preliminary data already being analyzed \citep{Kohn2019, Ghosh2020, Kern2020a, Kern2020b, Dillon2020, Thyagarajan2020}.

While these experiments have seen steady progress, one of their key limiting factors is the astrophysical foreground emission that dwarfs the 21\,cm cosmological signal by upwards of five orders of magnitude.
Coming predominately from non-thermal emission from diffuse structures associated with the Milky Way and extragalactic radio sources, these foregrounds create a delicate signal separation problem for 21\,cm experiments.
To meet this challenge, the past decade has seen significant progress in understanding how astrophysical foregrounds manifest in 21\,cm data \citep{Datta2010, Liu2011, Morales2012, Dillon2014, Thyagarajan2015a}.
Nonetheless, the foreground separation problem remains paramount to achieving a breakthrough in 21\,cm cosmology science.

An alternative in the short term is to avoid the foregrounds entirely, using the fact that while the foregrounds are spectrally smooth, the EoR is expected to be highly spectrally variant due to the inhomogeneities of the IGM.
Termed ``foreground avoidance,'' this may help current experiments in making an initial detection of the 21\,cm signal \citep[e.g.][]{Ali2015, Kerrigan2018}.
However, in throwing away parts of the data inherently contaminated by foregrounds, we lose a significant amount of information content: if we are able to model and suppress foregrounds, we stand to gain upwards a factor of five in sensitivity \citep{Pober2014}.
In particular, foreground emission is most prevalent at low line-of-sight Fourier $k$ modes, which are also the $k$ modes where the cosmological signal is expected to have the largest signal-to-noise.
Being able to suppress foregrounds below fiducial EoR levels could turn early, marginal evidence of the 21\,cm signal into a strong detection.

Fundamentally, one cannot get away with foreground avoidance entirely, as foreground modeling is generally needed for instrument calibration.
Considerable work has gone into constructing ever more precise catalogues of the low frequency radio sky for use in both calibration and foreground subtraction \citep[e.g.][]{Haslam1982, Oliveira2008, Zheng2016, Patil2017, Hurley-Walker2017}.
However, even the most precise foreground models to-date are unable to completely suppress foreground contamination.
As a consequence, other approaches to foreground subtraction have been developed that operate based on the assumption that foregrounds can be cast into a sparse basis, modeled and subtracted, without having to actually identify the source of the foreground emission in a sky map.
Some of these methods assume some prior knowledge of the foreground and signal, such as knowing their frequency covariances \citep[e.g.][]{Liu2011, Mertens2018} or assuming they can be described by a library of foreground and instrument models \citep{Tauscher2018}, while others make fewer up-front assumptions about the foregrounds (i.e. blind or semi-blind approaches) and seek to model their structure directy from the data \citep[e.g.][]{Harker2009, Chapman2013, Olivari2016, WOlz2017, Carucci2020}.
While blind techniques are favorable in that they make minimal assumptions about the foregrounds, these techniques can lead to oversubtraction of the EoR signal if the number of sparse foreground modes is not properly selected, which will be both an instrument and field dependent quantity that is not known a priori \citep{Ghosh2015}.
Bayesian approaches have also been explored, which aim to sample the joint posterior between the foreground and EoR power spectrum \citep{Zhang2016, Sims2019}.

Recently, a foreground separation method based on a Gaussian process (GP) model was proposed by \citet[][hereafter \citetalias{Mertens2018}]{Mertens2018}.
They claim that foreground modeling via Gaussian processes may be able to mitigate contaminants down to fiducial 21\,cm signal levels, and in particular, recover the low $k$ modes in the power spectrum for EoR measurements.
They test their method against realistic simulations and a wide variety of covariance models \citep[see also][]{Offringa2019}, and have recently demonstrated their method on real datasets from LOFAR \citep{Gehlot2018, Mertens2020} and HERA \citep{Ghosh2020}.
In particular, recent EoR power spectrum limits at $z-9.1$ by \citet[][hereafter \citetalias{Mertens2020}]{Mertens2020} expound upon the algorithm, showing that GPR foreground subtraction (GPR-FS) is inherently lossy and requires re-normalization.

In this work, we revisit the problem of Gaussian process-based modeling of 21\,cm foregrounds.
First, we cast GPR-FS into the framework of quadratic estimators (QE) of the power spectrum, putting the statistical properties of GPR-FS on stronger theoretical footing, and demonstrating how foreground separation and power spectrum estimation can be done simultaneously.
Incidentally, we show that GPR-FS is in fact closely related to the optimal quadratic estimator, widely used for galaxy and Cosmic Microwave Background (CMB) surveys \citep{Hamilton1997, Tegmark1997, Tegmark2002}, and more recently adapted for 21\,cm cosmology \citep{Liu2011}.
Our analysis reveals that GPR-FS distorts the window functions of the power spectrum at low $k$ modes, making it difficult to probe it below $k < 0.1\ {\rm Mpc}^{-1}$ without direct decorrelation.
Having cast GPR-FS into the QE framework, we next present novel extensions to GPR-FS that improve its performance and allow for it to handle certain real-world problems, like missing or nulled data due to radio frequency interference (RFI).

Lastly, we then look closely at recent upper limits on the 21\,cm power spectrum at $z=9.1$ set by LOFAR \citep{Mertens2020}, which relied on GPR foreground subtraction to access the low $k$ modes and improve the sensitivity of their limits.
Specifically, we show that their power spectrum normalization scheme is fairly sensitive to EoR signal loss when their adopted data covariance is misestimated, especially the EoR component of the covariance.
Indeed, \citetalias{Mertens2020} acknowledge the importance of a well-matched data covariance for their estimator, and trial a series of different covariances to select one that best fits their data, as well as employ a series of signal recovery tests to ensure their estimator is unbiased.
However, those signal recovery tests were confined to operate with the general covariances adopted by their analysis pipeline, whereas our analysis indicates that, given their choice of normalization, any deviation of the true EoR covariance from their assumption can lead to signal loss.
This has significant implications for recent astrophysical interpretations \citep[e.g.][]{Ghara2020, Mondal2020, Greig2020}, which ruled out models that do not conform to the covariance models adopted in \citetalias{Mertens2020}.
We also demonstrate that these issues can be largely mitigated by using the QE framework for implementing GPR-FS, where one can choose to trade power spectrum sensitivity for robustness against signal loss. 
Note that this work is not necessarily advocating for GPR-FS or inverse covariance weighting as the ideal foreground removal technique. Rather, we are simply re-evaluting the mathematical formalism of GPR-FS as a foreground subtraction method, showing its similarity to inverse covariance weighting, presenting some novel extensions to its standard form, and demonstrating its sensitivity to EoR signal loss when not properly normalized.

\citet{Hothi2020} recently compared the LOFAR GPR-FS technique against other blind separation techniques, showing that they agree at small $k$ modes and are outperformed by GPR-FS at high $k$ modes.
They also show that GPR-FS is less prone to overfitting than the blind separation methods studied, however, they do not explicitely investigate the concer raised by this work, which is the impact of misestimated covariance models for the GPR-FS method. Therefore, while it bolsters the results in \citetalias{Mertens2020}, the results of \citet{Hothi2020} address a different set of questions than the ones explored in this work.

While we focus our attention to experiments targeting the Cosmic Dawn and hydrogen reionization epochs, the techniques discussed in this work are equally applicable to 21\,cm experiments operating at higher frequencies targeting the post-reionization epoch \citep[e.g.][]{Chang2010, Masui2013, Shaw2015}, as well as intensity mapping experiments targeting other spectral lines \citep[e.g.][]{Cleary2016}.

\section{Data Modeling Formalism}
\label{sec:formalism}

In this section we describe the various components of the data vector for 21\,cm cosmology experiments, outlining how astrophysical foregrounds, the cosmological signal, and various instrumental contaminants appear in the data.
We then describe the formalism for Bayesian modeling of these components using a Gaussian process.
For more details on interferometric measurements see \citet{Hamaker1996, Smirnov2011}.
Also, for a comprehensive overview of Gaussian process modeling we refer the reader to \citet{Rasmussen2006}.

\subsection{Radio Inteferometric Measurements}
\label{sec:rime}

The fundamental observable for a radio interferometer is the visibility, $V_{jk}$, formed by the voltage correlation of two antennas $j$ and $k$, which is related to the specific intensity of the sky temperature via the measurement equation:
\begin{align}
\label{eq:me}
V_{jk}(\nu) = \int_{4\pi}d\Omega\ A(\hat{s}, \nu)I(\hat{s},\nu)e^{2\pi i\boldsymbol{b}_{jk}\cdot\boldsymbol{\hat{s}}\nu/c},
\end{align}
where $A$ is the direction and frequency dependent primary beam response of the antennas, $I$ is the sky temperature field, the exponential holds the interferometric fringe term, and the integral is over the full sky \citep{Thompson2017}.
Given a model of the sky and of the antenna primary beam response, the visibility is a complex-valued quantity that is uniquely specified by the baseline vector, $\boldsymbol{b}_{jk}$.

The form of \autoref{eq:me} tells us something about the amount of \emph{spectral structure} that is contained within the visibility at any given time.
Low-frequency radio foreground emission is generally regarded to be spectrally smooth, as it is thought to be predominately generated by non-thermal synchrotron emission with a characteristic spectral slope of roughly $\nu^{-2.2}$ \citep{Condon1992}.
If we were to Fourier transform this across frequency, we would find its power confined to low Fourier modes, indicative of its inherent smoothnesss.
However, in \autoref{eq:me}, we can see that this is not the only frequency dependent term that factors into the visibility.
In particular, the intrinsic fringe response of the interferometer has its own frequency dependence, which scales with the projection of the source location on the sky onto the baseline vector.
To make this more evident, we can recast the fringe term in \autoref{eq:me} in terms of the time delay of the source signal between the two antennas,
\begin{align}
\label{eq:delay}
e^{2\pi i\boldsymbol{b}_{jk}\cdot\hat{s}\nu/c} = e^{2\pi i\tau_{jk}(\hat{s})\nu},
\end{align}
where $[\tau]$ has units of seconds and is the Fourier dual of frequency.
Here we see that the fringe term can be thought of as a delta function in delay space centered at $\tau_{jk}(\hat{s})$, where its role is to convolve the intrinsically compact response of $I$ and spread its power to larger Fourier modes.
This is, in fact, the phenomenon also known as the foreground wedge \citep{Datta2010, Morales2012, Parsons2012a, Liu2014a}.
However, we can see that the fringe term has a maximum achievable delay, occuring when the source of radiation is incident from the horizon, in which case
\begin{align}
\tau_{jk}^{\rm horizon} = b_{jk}/c.
\end{align}
This tells us that foreground structure in the visibilities should be confined to delays $|\tau_{jk}| < \tau_{jk}^{\rm horizon}$, which is useful for constructing a data-based foreground model.
In practice, the other terms in \autoref{eq:me} like the beam and the intrinsic gain response also contribute to additional spectral structure \citep{Barry2016, Ewall-Wice2017, Dillon2018, Byrne2019, Kern2020b, Joseph2020}.
However, this work will for the time being sidestep the problem of calibration and beam modeling (deferred to future work), and assume that these have been mitigated sufficiently by an analysis pipeline.

If one is working with the \emph{gridded visibilities}, which grids and co-adds the visibilities in fixed bins in $\boldsymbol{u}=\boldsymbol{b}\nu/c$ across different frequencies, then the frequency response of the interferometric fringe term has technically been taken out, and the foreground wedge is no longer an issue.
However, in practice, this gridding has never quite been achieved to the necessary precision, and residual wedge-like artifacts are generally still observed even in the gridded visibilities \citep{Morales2019}.

\subsection{Gaussian Process Regression}
\label{sec:gpr}

Here we review Gaussian process regression, drawing from \citet{Rasmussen2006} for the general formalism, and \citetalias{Mertens2018} for applications to 21\,cm foreground subtraction.

Let our dependent data be described as a column vector, $\d$, containing the instrumental measurements at each frequency, and let the column vector $\boldsymbol{\nu}$ be the array of frequencies, such that
\begin{align}
\label{eq:dvec}
\d = \begin{bmatrix}d_1\\ d_2\\ \vdots\\ d_n\end{bmatrix}; \boldsymbol{\nu} = \begin{bmatrix}\nu_1\\ \nu_2\\ \vdots\\ \nu_n\end{bmatrix}.
\end{align}
Usually this data vector is a gridded or ungridded interferometric visibility, although one could also take the data vector to be the frequency response of a pixel from a sky map.
Multiple observations can be incorporated by expanding our data vector as an $n\times m$ array, where $m$ is the number of integrations in the data.
Multiple visibilities (i.e. the instrument response from different baseline vectors) can similarly be accounted for by appending along this axis.
Nominally, our data vector consists of the measured foreground ($f$), EoR signal ($e$), and thermal noise ($n$),
\begin{align}
\label{eq:data_vec}
\d = \f + \e + \n,
\end{align}
which we assume to be statistically uncorrelated.
Thus the covariance of the data is given by
\begin{align}
\label{eq:data_cov}
\C = \langle\d\d^\dagger\rangle = \Cfg + \Cto + \Cn,
\end{align}
where $\langle\rangle$ denotes an ensemble average.
Assuming our data vector is Gaussian distributed, we can model its probability distribution as
\begin{align}
\d\sim\mathcal{N}\left(m(\bnu), K(\bnu, \bnu)\right),
\end{align}
where $m$ is an underlying mean function of our model and $K$ is its covariance function (or kernel).
In other words, we assume that our data vector is a discrete realization of a class of continuous fields drawn from a Gaussian distribution in function space, also known as a Gaussian process.
Note that for a point in the $uv$ plane with non-zero magnitude, the complex visibility has an ensemble average of zero, thus we assume an identically zero mean function for modeling the complex visibilities.

Given our measurements, $\d$, we can write down the joint distribution of the Gaussian process at a series of other points in space $\bnu^\prime$ as
\begin{align}
\label{eq:joint_gp}
\begin{bmatrix}\d \\ \d^\prime\end{bmatrix} \sim \mathcal{N}\left(\begin{bmatrix}0\\ 0\end{bmatrix},
\begin{bmatrix}K(\bnu,\bnu) & K(\bnu,\bnu^\prime)\\K(\bnu^\prime, \bnu) & K(\bnu^\prime,\bnu^\prime)\end{bmatrix}\right).
\end{align}
In general GP applications, we are usually interested in learning something about the Gaussian process at new points in space given our knowledge of the measured data vector.
To get this, we condition the Gaussian process on our measurements, yielding a probability distribution on $\d^\prime$
\begin{align}
\d^\prime|\d \sim \mathcal{N}\left(\Exp[\d^\prime], \Cov[\d^\prime]\right),
\end{align}
where $\Exp[]$ is the expectation value, $\Cov[]$ is the covariance, and
\begin{align}
\label{eq:gp_conditional}
\Exp[\d^\prime] &= K(\nu^\prime,\nu)K(\nu,\nu)^{-1}\d \\
\Cov[\d^\prime] &= K(\nu^\prime,\nu^\prime) - K(\nu^\prime, \nu)K(\nu, \nu)^{-1}K(\nu, \nu^\prime).
\end{align}

For the case of foreground modeling in interferometric visibility data, what we want to extract is an estimate of our foreground model $f^\prime$ given our measurements $d$ (generally at the same frequencies $\bnu$).
To do this, we pick a model for the covariance of each term in the data, $K = \Kfg + \Kto + \Kn$, and plug them into \autoref{eq:joint_gp}.
This gives us the joint distribution on our data and estimated foreground model,
\begin{align}
\begin{bmatrix}\d \\ \f^\prime\end{bmatrix} \sim \mathcal{N}\left(\begin{bmatrix}0\\ 0\end{bmatrix},
\begin{bmatrix}\Kfg+\Kto+\Kn & \Kfg\\ \Kfg & \Kfg\end{bmatrix}\right),
\end{align}
where the off-diagonal terms in the joint density covariance matrix simplify to $\Kfg$ because the foreground vector is the only data term that is correlated between $\f^\prime$ and $\d$.
Using \autoref{eq:gp_conditional}, we then find that
\begin{align}
\label{eq:fg_conditional}
\Exp[\f^\prime] &= \Kfg [\Kfg+\Kto+\Kn]^{-1}\d \\
\label{eq:fg_cov_conditional}
\Cov[\f^\prime] &= \Kfg - \Kfg[\Kfg+\Kto+\Kn]^{-1}\Kfg.
\end{align}
For foreground removal applications, we are interested in forming the data residual vector,
\begin{align}
\label{eq:res}
\r = \d - \Exp[\f^\prime],
\end{align}
which we hope subtracts the foregrounds and reveals the 21\,cm signal.
In this work, we will use the term ``residual'' to mean exclusively the vector $\r$.
Note that there is a difference between the covariance kernel we choose to model a specific term in the data vector, (e.g. $K_{\rm fg}$) and the actual covariance of that term (e.g. $\Cfg$), the latter of which is generally never known exactly.
Ideally we want these to be as close as possible, but they are not by construction equivalent.
In \autoref{sec:gprfs} we assume that the true covariance is known, or at least is known to good enough approximation, and in \autoref{sec:lofar} we demonstrate the impact on the power spectrum when this assumption fails.

The covariance of the various terms in our data are each assigned their own covariance, which are themselves dependent on a set of \emph{hyperparameters} that can either be fixed with prior knowledge or regressed for directly from the data.
A common covariance model (or kernel) used in Gaussian Process modeling is the Mat\'{e}rn kernel,
\begin{align}
\label{eq:matern}
\Km(\nu, \nu^\prime) = \sigma^2\frac{2^{1-\eta}}{\Gamma(\eta)}\left(\sqrt{2\eta}\frac{|\nu-\nu^\prime|}{\ell}\right)^\eta K_\eta\left(\sqrt{2\eta}\frac{|\nu-\nu^\prime|}{\ell}\right),
\end{align}
where $\sigma^2$ is its variance parameter, $\ell$ is its lengthscale parameter, $\eta$ is its spectral parameter (together being the three hyperparameters of the Matern kernel), $\Gamma$ is the gamma function, and $K_\eta$ is the modified Bessel function of the second kind \citep{Rasmussen2006}.
The variance parameter sets the overall amplitude of the signal, while the length scale parameter sets the typical scale of correlations (in our case across frequency).
The spectral parameter $\eta$ sets the overall ``smoothness'' of the signal.
For particular choices of $\eta$, the Mat\'{e}rn kernel dramatically simplifies, most notably to a radial basis function (or squared exponential function) when $\eta\rightarrow\infty$.
For all other half-integer $\eta$, the Mat\'{e}rn kernel can be written as the product of an exponential and polynomial, with $\eta=1/2$ resulting in a purely exponential function.
Most of the covariance models used in this work are related to the Mat\'{e}rn kernel.

In \autoref{sec:gpr_windows}, for example, we model the foreground component of our data with a $\eta\rightarrow\infty$ Mat\'{e}rn kernel (or a radial basis function or a Gaussian function),
\begin{align}
\label{label:fg_kernel}
\Kfg(\nu, \nu^\prime) = \sigmafg^2\exp\left(-\frac{(\nu-\nu^\prime)^2}{2\ellfg^2}\right),
\end{align}
which is motivated by the fact that foregrounds, although imparted with some spectral structure by the instrument, are thought to be largely spectrally smooth.
Recall we outlined the rough amount of foreground spectral structure we expect to see in a visibility depending on its baseline length and what frequency scale that translates to (\autoref{sec:rime}).
We therefore have a good guess for what the foreground covariance hyperparameters should be, although we will return to how we actually refine these estimates in a data-driven manner.
\citetalias{Mertens2018} go one step further and break down the foreground covariance into two separable covariances, one for modeling broad scale foreground fluctuations and one for modeling intermediate scale, instrument-induced fluctuations, both using a form of Mat\'{e}rn kernel.

The 21\,cm signal is expected to be much more spectrally variant and is not well modeled by a radial basis function.
Similar to \citetalias{Mertens2018}, we adopt an exponential function for the 21\,cm signal,
\begin{align}
\label{label:eor_kernel}
\Kto(\nu, \nu^\prime) = \sigmato^2\exp\left(-\frac{|\nu-\nu^\prime|}{\ellto}\right).
\end{align}
\citetalias{Mertens2018} found that an exponential distribution roughly agrees with the frequency covariance of some of the existing fiducial 21\,cm EoR models.
However, this agreement does not extend throughout the currently allowable EoR parameter space (something we return to in \autoref{sec:bias_correction_math}).

Lastly, we use a simple scalar matrix to model the noise, which is uncorrelated from frequency channel-to-channel,
\begin{align}
\Kn(\nu, \nu^\prime) = \sigman^2\delta_{\nu\nu^\prime},
\end{align}
where $\delta_{\nu\nu^\prime}$ is the Kronecker delta, which is one if $\nu=\nu^\prime$ and zero otherwise.

Note that we've chosen to model our covariance functions as stationary: in other words, they are independent of where in $\nu$ they are evaluated and only depend on $\nu-\nu^\prime$.
However, this is not strictly true for some of the terms in our data, as we saw in \autoref{sec:rime}.
For example, we know that low-frequency foregrounds become brighter at lower frequencies.
Similarly, thermal noise is also non-stationary, rising at low frequencies due to the increase in sky temperature.
For transit telescopes like HERA, these terms are also \emph{time dependent}, as the foreground and noise statistics will change depending on what kinds of structures are transiting the field of view.
For this work, we concern ourself with a fairly narrow spectral window and only consider a narrow set of times, making these considerations less important.
However, for wide-band and multi-integration foreground modeling, these effects will likely need to be taken into account.

\subsubsection{Hyperparameter Optimization}
\label{sec:optimization}
The kernel hyperparameters can be selected using Bayesian model selection to determine the values that best fit the data.
This is done by maximizing the marginal likelihood of the data, or the Bayesian evidence, which is the integral of the data likelihood times the prior on our function values,
\begin{align}
p(\d|\bnu, \btheta) = \int \mathcal{L}(\d|\g,\bnu,\btheta)p(\g|\bnu,\btheta)d\g,
\end{align}
where $\mathcal{L}$ is the likelihood, $\g$ is another vector spanning the space of our data and $\btheta$ is a vector containing our kernel hyperparameters.
This Gaussian integral can be done analytically, and is given up to a constant as
\begin{align}
\label{eq:ml}
\log p(\d|\bnu,\btheta) \propto -\frac{1}{2}\d^\dagger K(\bnu,\bnu|\btheta)^{-1}\d - \frac{1}{2}\log{\rm det}|K(\bnu,\bnu|\btheta)|,
\end{align}
where recall $K$ is the full covariance model of our data \citep{Rasmussen2006}.
The log posterior distribution on our hyperparameters is then proportional to the log marginal likelihood plus the log prior: $\log p(\d|\bnu,\btheta) + \log p(\btheta)$.
We use a Markov chain Monte Carlo (MCMC) sampler from the \texttt{emcee} package \citep{Foreman-Mackey2013, Goodman2010} to explore the joint hyperparameter posterior distribution.
To evaluate the Gaussian process equations throughout this work, we use the \texttt{scikit-learn}\footnote{\url{https://scikit-learn.org/stable/}} package's optimized implementation \citep{Pedregosa2012}, which themselves are derived from \citet{Rasmussen2006}.

\section{GPR Foreground Subtraction in the Quadratic Estimator Framework}
\label{sec:gprfs}

In this section we outline how GPR foreground subtraction (GPR-FS) can be incorporated into the quadratic estimator (QE) formalism.
We start by giving a brief review of quadratic estimators, and then demonstrate GPR-FS.
Principally, we show that this is closely related to the inverse covariance estimator outlined in \citet{Liu2011}.
We also discuss some variations to GPR-FS that allow for estimation of the foreground component (which is often useful for null testing and data quality management), in addition to handling the issue of missing data (e.g. due to RFI flags).
Throughout this section we sidestep the problem of estimating the data covariance, and assume it is known.
Although previous works have demonstrated how one can estimate the data covariance empirically \citep{Dillon2014}, if not done carefully, this has also shown to cause inadvertent suppression of the cosmological signal \citep[e.g.][]{Switzer2014, Cheng2018}.

\subsection{Quadratic Estimators of the Power Spectrum}
\label{sec:qe}

The 21\,cm power spectrum is the square of the three dimensional Fourier transform of the temperature field,
\begin{align}
\langle\widetilde{T}(\boldsymbol{k})\widetilde{T}^\ast(\boldsymbol{k}^\prime)\rangle = (2\pi)^3P(k)\delta(\boldsymbol{k}-\boldsymbol{k}^\prime),
\end{align}
where $\widetilde{T}$ is the Fourier transformed temperature field, and $\delta$ is the Dirac-delta function.
Noting that the interferometric visibilities are the transverse Fourier transform of the sky (under the flat sky approximation), an estimate of the power spectrum can be made by simply Fourier transforming the gridded visibilities across frequency.
Under certain approximations, one can also estimate the power spectrum by Fourier transforming the ungridded visibilities, known as the delay spectrum estimator \citep{Parsons2012a}, which is being used by highly redundant arrays like HERA.
In this case, the power spectrum is defined as
\begin{align}
\label{eq:dspec}
P(k_\perp, k_\para) = \frac{X^2Y}{\Omega_pB}\langle\widetilde{V}(\boldsymbol{b}/\lambda, \tau)\widetilde{V}^\ast(\boldsymbol{b}/\lambda, \tau)\rangle,
\end{align}
where $X$ and $Y$ are scalars mapping angles on the sky to transverse cosmological distances and frequencies to line-of-sight cosmological distances, $\Omega_p$ is the integral of the primary beam, $B$ is the frequency bandwidth in the Fourier transform, and $\widetilde{V}$ is the Fourier transformed visibilities.
The cosmological wavevectors are related to the ``telescope units'' via
\begin{align}
\label{eq:kvecs}
k_\perp &= \frac{2\pi\tau}{X} \\
k_\para  &= \frac{2\pi}{Y}\frac{b}{\lambda},
\end{align}
where $X = c(1+z)^2\nu_{21}^{-1}H(z)^{-1}$, $Y=D(z)$, $\nu_{21}=1.420$ GHz, $H(z)$ is the Hubble parameter, $\lambda$ is the average observing wavelength, and $D(z)$ is the transverse comoving distance \citep{Parsons2014}.
In this work we adopt a $\Lambda$CDM cosmology with parameters derived from the Planck 2015 analysis \citep{Planck2016}.

To be more statistically rigorous, we can cast the delay spectrum technique into the quadratic estimator framework \citep{Hamilton1997, Tegmark1997, Liu2014a}.
We start by approximating the continuous power spectrum as a piecewise function in sufficiently narrow bins in $k$.
The $\alpha$ bin's amplitude, $p_\alpha$, is referred to as its band power, and its reconstruction from the data is the ultimate goal of our estimator.
Fundamentally, we assume that the covariance of the data in frequency space can be related to the band powers as
\begin{align}
\label{eq:bandpowers}
\C = \N + \Cfg + \Cto = \N + \Cfg + \sum_\alpha p_\alpha \C_{,\alpha},
\end{align}
where $\N$, $\Cfg$, and $\Cto$ are the covariance of our data terms in frequency space, $p_\alpha$ are the band powers, and $\C_{,\alpha} = \partial\Cto/\partial p_\alpha$ is the derivative of the EoR covariance with respect to the band powers.
We are free to define this matrix however we see fit (thereby setting the definition of the band power itself), so long as \autoref{eq:bandpowers} is satisfied.
In almost all practical cases, we generally set $\C_{,\alpha}$ to be the discrete Fourier transform (DFT) operator.
A DFT basis for band powers is convenient because of its orthonormality, as well as its straightforward interpretation as the descretized power spectrum.
Although one can fold more data manipulation steps into $\C_{,\alpha}$ \citep[e.g.][]{Liu2011}, in this work we choose to define $\C_{,\alpha}$ as simply a Fourier transform operator, which can be decompsed as
\begin{align}
\label{eq:C_alpha}
\C_{,\alpha} = \c_\alpha\c_\alpha^\dagger,
\end{align}
where $\c_\alpha$ is a length-$N$ DFT column vector whose elements are defined as
\begin{align}
\label{eq:dft}
c_\alpha^m = \frac{1}{N}e^{-2\pi i\alpha m/N},
\end{align}
with $N$ representing the number of frequency bins.
Thus its main role is to map the data vectors on either side of it from frequency to the Fourier domain.

A quadratic estimate of the band powers can be formed from our data vectors as
\begin{align}
\label{eq:qe}
\hat{p}_\alpha = \d^\dagger\E_\alpha\d - \hat{b}_\alpha,
\end{align}
where $\d$ is our data vector, $\hat{p}_\alpha$ is our estimate of the $\alpha$ band power, $\E$ is a family of matrices to-be-described, and $\hat{b}_\alpha$ is our estimate of the residual bias term.
Note that while the band powers are quadratic with respect to the data, the quadratic estimator is a linear mapping between data pairs and the power spectrum.
Taking the expectation value of \autoref{eq:qe} and substituting in \autoref{eq:bandpowers}, we find that
\begin{align}
\label{eq:window_funcs}
\langle\hat{p}_\alpha\rangle = \sum_\beta\tr[\C_{,\beta}\E_\alpha]p_{\beta} + \tr[(\N + \Cfg)\E_\alpha] - \hat{b}_\alpha.
\end{align}
\autoref{eq:window_funcs} tells us that for our estimator to be unbiased we set $\hat{b}_\alpha = \tr[(\N+\Cfg)\E_\alpha]$.\footnote{This is discussed in more detail in \autoref{sec:bias_correction_math}.}
Furthermore, it shows us that the estimated band powers $\hat{p}_\alpha$ are a linear combination of the true band powers $p_\alpha$ dictated by a transfer matrix, which we will denote as $\W$, commonly known as the \emph{window function},
\begin{align}
\label{eq:WF}
\langle\hat{\p}\rangle = \W\p.
\end{align}
The covariance of our band powers subject to the covariance of our input data can be written as
\begin{align}
\label{eq:qe_C}
\Sigma_{\alpha\beta} = \langle \hat{p}_\alpha\hat{p}_\beta\rangle - \langle \hat{p}_\alpha\rangle\langle \hat{p}_\beta\rangle = 2\tr[\C\E_\alpha\C\E_\beta].
\end{align}
Note that the covariance of the band powers, $\bSigma$, is defined in the Fourier domain, whereas the covariance of the data, $\C$, is defined in the frequency domain.
The band powers, their window functions, and their covariance matrix completes our full statistical accounting of the EoR power spectrum for a Gaussian random field \citep{Liu2014a}.

\begin{figure*}
\centering
\includegraphics[width=\linewidth]{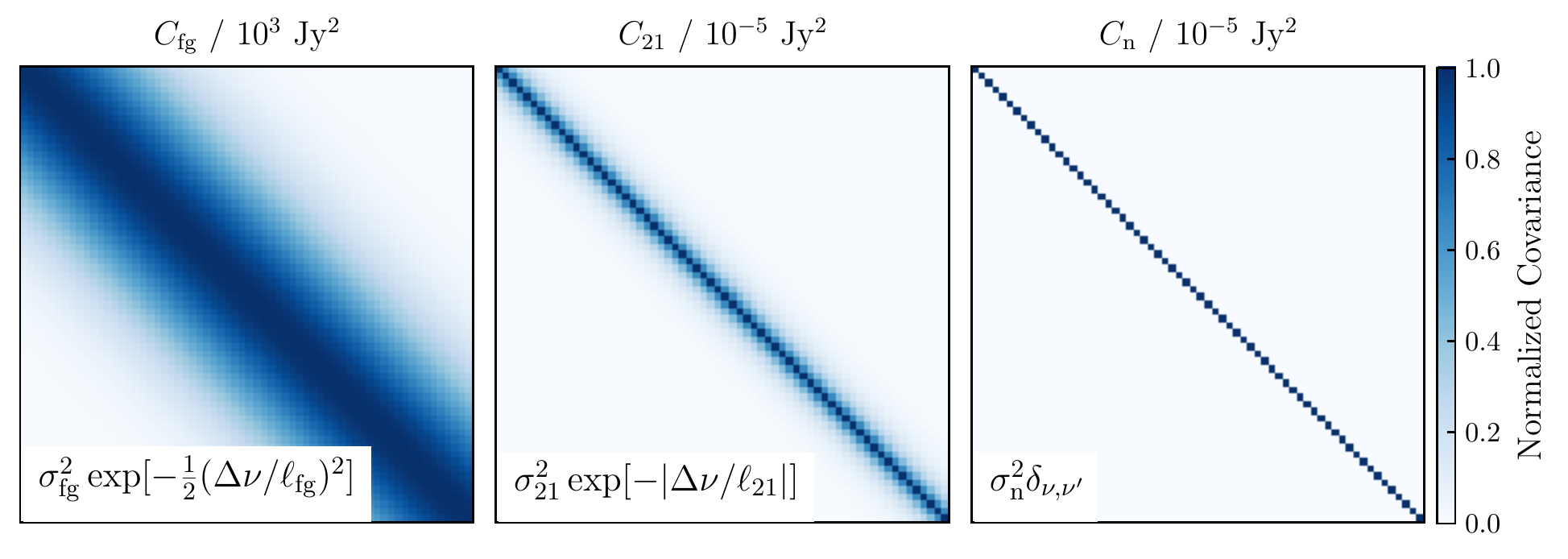}
\caption{The covariance of the foregrounds, EoR, and noise components of our analytic data simulations in \autoref{sec:gprfs}.
Parameters for the foreground covariance ($\sigmafg^2=10^2\ {\rm Jy}^2$, $\ellfg=4$ MHz) are selected to be roughly characteristic of HERA data.
The EoR covariance parameters ($\sigmato^2=10^{-5}\ {\rm Jy}^2$, $\ellfg=0.75$ MHz) are selected to be roughly in line with fiducial theoretical expectations \citep{Mertens2018}, and the noise is artificially selected depending on the amount of integration assumed for the data.}
\label{fig:data_covariance}
\end{figure*}

How do we actually compute these quantities?
The $\E_\alpha$ matrix can be decomposed as
\begin{align}
\label{eq:qe_E}
\E_\alpha = \frac{1}{2}\sum_\beta \M_{\alpha\beta} \R^\dagger \C_{,\beta} \R,
\end{align}
where $\M$ is a normalization matrix that ensures the estimated band powers are properly normalized, and $\R$ is an arbitrary data-weighting matrix, both of which the data analyst is free to choose subject only to the constraint that our band powers are properly normalized, or that $\sum_\beta\W_{\alpha\beta}=1$.
Here we also see the main role of $\C_{,\alpha}$ as a Fourier transform operator for the weighted data vectors on either side of it.
Given a choice of $\C_{,\alpha}$ and $\R$, we can form the quantity
\begin{align}
\label{eq:H_mat}
H_{\alpha\beta} = \frac{1}{2}\tr[\R^\dagger\C_{,\alpha}\R\C_{,\beta}],
\end{align}
which, according to the first term in \autoref{eq:window_funcs}, conveniently lets us re-express the window functions as
\begin{align}
\label{eq:window_func}
\W = \M \H.
\end{align}

The commonly adopted $\R$ matrix in the literature is $\R = \C^{-1}$, or the inverse data covariance matrix.
Given this weighting matrix and assuming a diagonal normalization matrix of $M_{\alpha\beta} = \delta^k_{\alpha\beta}\left(H_{\alpha\beta}\right)^{-1}$ where $\delta^k$ is the Kronecker delta, it can be shown that we recover the optimal, minimum variance estimator of the power spectrum \citep{Hamilton1997, Tegmark1997}.
In such a limit, we get that $\H = \F$ and $\boldsymbol{\Sigma}_{\alpha\alpha} = (\F_{\alpha\alpha})^{-1}$, where $\F$ is the band power Fisher matrix.
Alternative choices of the normalization matrix lead to other desirable properties, such as $\M=\H^{-1}$ leading to window functions that are diagonal in $k$ space, and $\M=\H^{-1/2}$ leading to a band power covariance that is uncorrelated between $k$ modes \citep{Tegmark2002}.
While $\M=\H^{-1}$ deconvolves the window function, it does so at the expense of larger errorbars.
We therefore only use the minimum variance normalization (referred to as $\M\propto\I$) and the uncorrelated covariance normalization (referred to as $\M=\H^{-1/2}$) in this work.
Regardless of the chosen normalization matrix, we refer to the optimal quadratic estimator as one that adopts $\R_{\rm OQE} = \C^{-1}$, which is also the ``inverse variance weighted foreground removal estimator'' described in \citet{Liu2011}.

To cast Gaussian process foreground subtraction into the quadratic estimator framework, we recognize that \autoref{eq:res} is simply a linear operation on the data vector and can be equivalently expressed as
\begin{align}
\label{eq:gprfs}
\r = (\I - \Kfg[\Kfg + \Kto + \Kn]^{-1})\d  = \R_{\operatorname{GPR-FS}}\d.
\end{align}
A quadratic estimator that enacts a GPR foreground subtraction as part of its weighting matrix (referred to as GPR-FS) is therefore achieved by adopting the above $\R$ matrix.
Furthermore, variations about this technique are possible by dotting this matrix into other subsequent operators, such as additional, post-foreground removal weighting matrices, which we discuss in the following sections.

After squaring the visibilities to form the power spectrum, we can further average them together assuming isotropy and homogeneity of the cosmological signal by carving spherical shells in $\bk$ space across lines of constant $|\bk|$.
Assuming the mapping between the band powers of the true spherical power spectrum to the discretized band powers of the estimated per-visibility power spectrum can be described by a design matrix and an error term,
\begin{align}
\hat{\p}_{\rm vis} = \A\p_{\rm sph} + \boldsymbol{\epsilon},
\end{align}
then the optimal compression of the band powers onto a spherical basis \citep{Dillon2014} is given as
\begin{align}
\hat{\p}_{\rm sph} = \left[\A^T\bSigma_{\rm vis}^{-1}\A\right]^{-1}\A^T\bSigma_{\rm vis}^{-1}\hat{\p}_{\rm vis},
\end{align}
with a covariance matrix
\begin{align}
\bSigma_{\rm sph} = \left[\A^T\bSigma_{\rm vis}\A\right]^{-1}
\end{align}
and a window function matrix
\begin{align}
\label{eq:sph_window}
\W_{\rm sph} = \left[\A^T\bSigma_{\rm vis}^{-1}\A\right]^{-1}\A^T\bSigma_{\rm vis}^{-1}\W_{\rm vis}\A.
\end{align}
For each estimated band power, $\hat{p}_\alpha$, its vertical error bar is given as $\sqrt{\Sigma_{\alpha\alpha}}$, even though there may be non-zero off-diagonal components for the $\alpha$ band power.
If we regard the window function as a probability distribution, we can estimate the central Fourier bin of the band power, $\hat{k}_\alpha$, as the window function's 50th percentile.
The difference between this estimate, $\hat{k}$, and the $k$ mode we would have naively measured with a simple ``Fourier transform and square'' estimator is one of the key takeaways from this section.
Furthermore, the horizontal errorbars of the band power can be estimated as the window function's 16th and 84th percentile, or its 68\% confidence interval in the Gaussian limit.

\subsection{The Relationship Between GPR-FS and OQE}
\label{sec:gpr_oqe}

Here we show that the GPR foreground subtracted quadratic estimator (GPR-FS) is closely related to the optimal quadratic estimator, and is in fact identical to OQE when dotting the former with an additional inverse covariance weighting matrix.
For the time being, we will assume we know the true covariance of the data and will adopt terms like $\Cfg$ instead of $\Kfg$ in our equations.
We can show the relationship between these estimators by using the Woodbury matrix identity, which states that for an invertible matrix $\A$ and for a matrix $\B$ that can be decomposed as $\B = \U\S\V^\dagger$,
\begin{align}
[\A + \U\S\V^\dagger]^{-1} = \A^{-1} - \A^{-1}\U[\S^{-1} + \V^\dagger\A^{-1}\U]^{-1}\V^\dagger\A^{-1}.
\end{align}
In our case, both $\A$ and $\B$ are taken to be Hermitian covariance matrices, implying that $\U$ and $\V$ are unitary matrices subject to $\U\V^\dagger = \I$, with $\S$ being a diagonal eigenvalue matrix.
Re-arranging, we find that
\begin{align}
[\A + \U\S\V^\dagger]^{-1} &= \A^{-1}(\I - \U[\A\U\S^{-1} + \U]^{-1}) \nonumber \\ 
&=\A^{-1}(\I - \U\S\V^\dagger[\A + \U\S\V^\dagger]^{-1})
\end{align}
where in the first equality we pulled $\V^\dagger\A^{-1}$ into the bracketed inverse and used the fact that $[\V^\dagger]^{-1} = \U$, and in the second equality formed the quantity $\U = \U\S\V^\dagger[\S\V^\dagger]^{-1}$ and pulled part of it into the bracketed inverse.
Substituting in $\A = \Cn + \Cto$ and $\B = \Cfg$, we get that
\begin{align}
[\Cn + \Cto + \Cfg]^{-1} &= [\Cn + \Cto]^{-1}(\I - \Cfg[\Cn + \Cto +  \Cfg]^{-1}) \nonumber \\
\R_{\rm OQE} &= [\Cn + \Cto]^{-1}\R_{\operatorname{GPR-FS}}
\end{align}
which tells us that a GPR-FS matrix dotted with an additional inverse noise and EoR covariance matrix is equal to the inverse of the full data covariance matrix.
Given that the statistical properties of a quadratic estimator are uniquely defined by the choice of $\R$ matrix (holding our choice of $\C_{,\alpha}$ constant), we conclude that the optimal quadratic estimator and the inverse variance weighted GPR-FS estimator are actually identical power spectrum estimators.
This is an interesting and non-intuitive result that helps us understand the properties of both of these methods.
First, we see that the inverse covariance foreground separation method of \citet{Liu2011}--the OQE of the 21\,cm signal in the presence of foregrounds--can in fact be re-cast as a Gaussian process, with \autoref{eq:fg_conditional} yielding its constrained foreground estimate.
Second, we see that the window function and band power covariance matrices for GPR foreground subtraction can be easily derived, and are in fact similar to those for the optimal estimator.
Additionally, this connects our understanding of the properties of inverse covariance weighting outlined in \citet{Liu2011} (namely that of $\C^{-1}$ acting as a filtering matrix that needs to be re-normalized in order to make an unbiased measurement of the 21\,cm signal) to GPR foreground subtraction, which we will revisit in \autoref{sec:lofar}.

\begin{figure*}
\centering
\includegraphics[width=\linewidth]{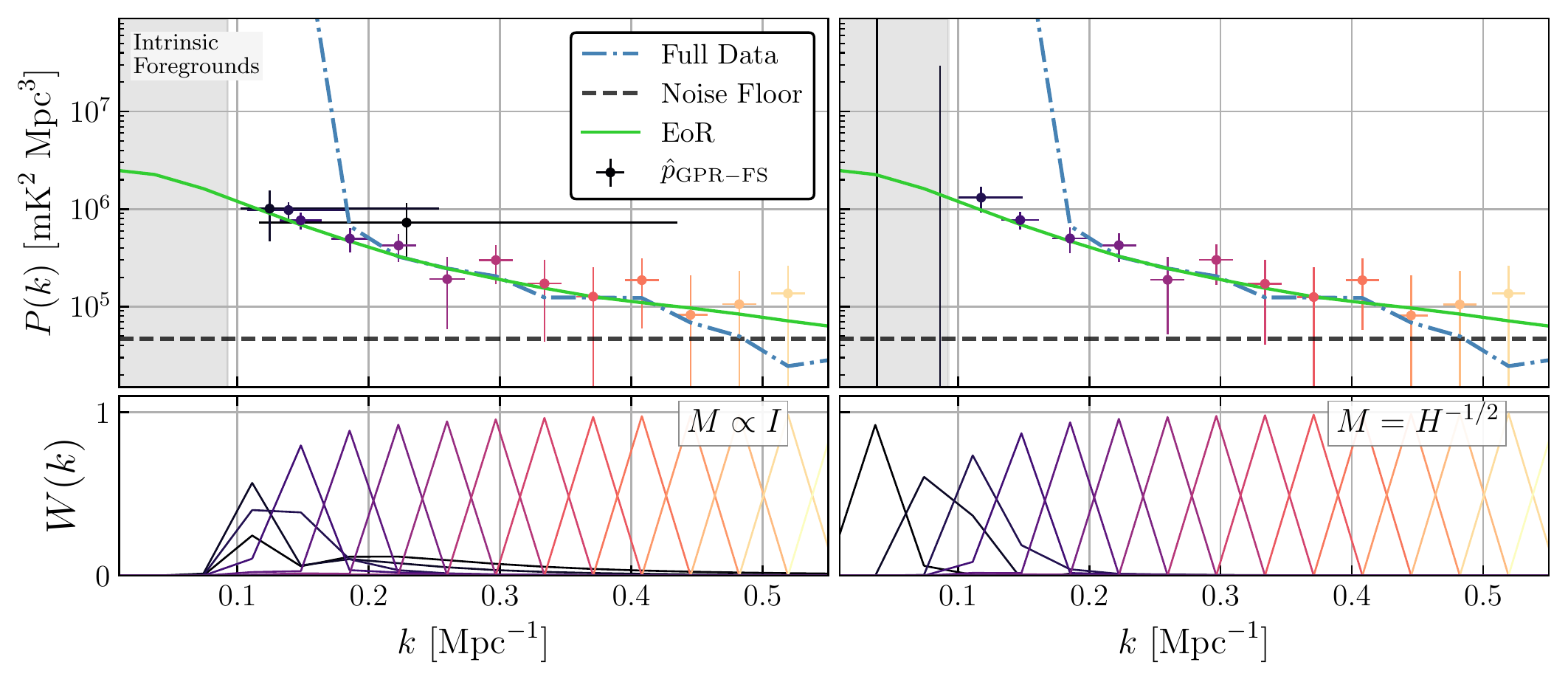}
\caption{GPR foreground subtraction and power spectrum estimation in a low-noise scenario for two different normalization schemes: $\M\propto\I$ (left panel) and $\M=\H^{-1/2}$ (right panel).
In this low-noise scenario, the intrinsic EoR is measured at an SNR $\sim$ 10 at $k\sim0.2\ h\ {\rm Mpc}^{-1}$.
We plot the full dataset including foregrounds, EoR, and noise (blue dot-dashed), the intrinsic EoR component (green), the noise floor (black-dashed), and the recovered band powers (points), which are centered at the $k$ associated with the 50\% quantile of the window function.
Vertical errorbars are the $1\sigma$ uncertainty, while the horizontal errorbars show the 16\% and 84\% quantile of the window function.
We see that the $\M\propto\I$ normalization scheme is unable to probe the power spectrum for $k\lesssim 0.13\ h\ {\rm Mpc}^{-1}$, even though its vertical errorbars are smaller than that of the $\M=\H^{-1/2}$ normalization scheme.
}
\label{fig:gpr_window_low_noise}
\end{figure*}

Lastly, we note that GPR foreground subtraction need not only apply to foregrounds.
By simply substituting $\Kfg$ with the sum of any number of (uncorrelated) covariances describing terms in the data that we would like to subtract, we can construct a GPR-based signal separation estimator that is near equivalent to what the optimal quadratic estimator would achieve.

\begin{figure*}
\centering
\includegraphics[width=\linewidth]{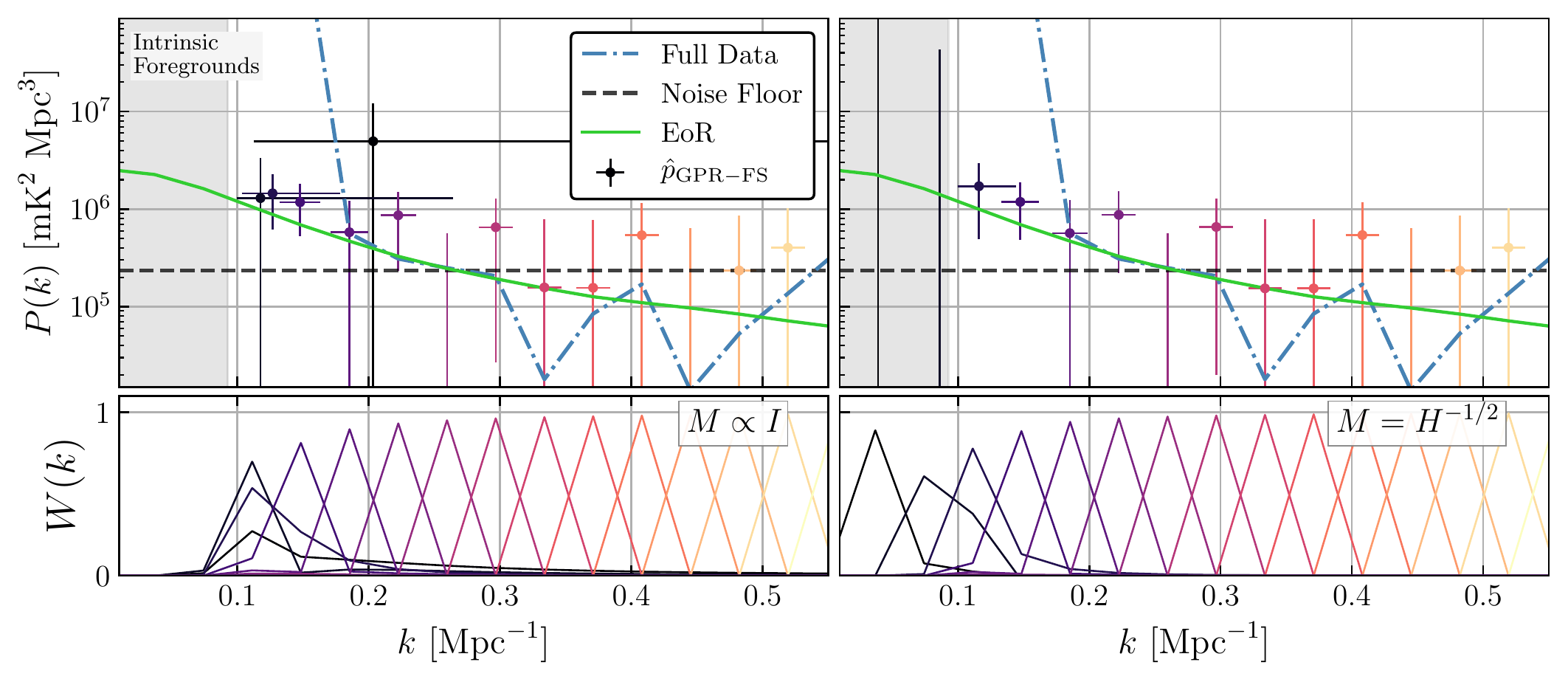}
\caption{The same as \autoref{fig:gpr_window_low_noise} but in a moderate-noise scenario, where the EoR appears with SNR$\sim$1 at $k\sim0.2\ h\ {\rm Mpc}^{-1}$.
This scenario is more realistic for a first-detection of the 21\,cm signal by HERA or the SKA.
GPR-FS with $\M\propto\I$ normalization is still unable to probe the power spectrum at low $k$ in this scenario, although the window functions at low $k$ are marginally better behaved.}
\label{fig:gpr_window_med_noise}
\end{figure*}

\begin{figure*}
\centering
\includegraphics[width=\linewidth]{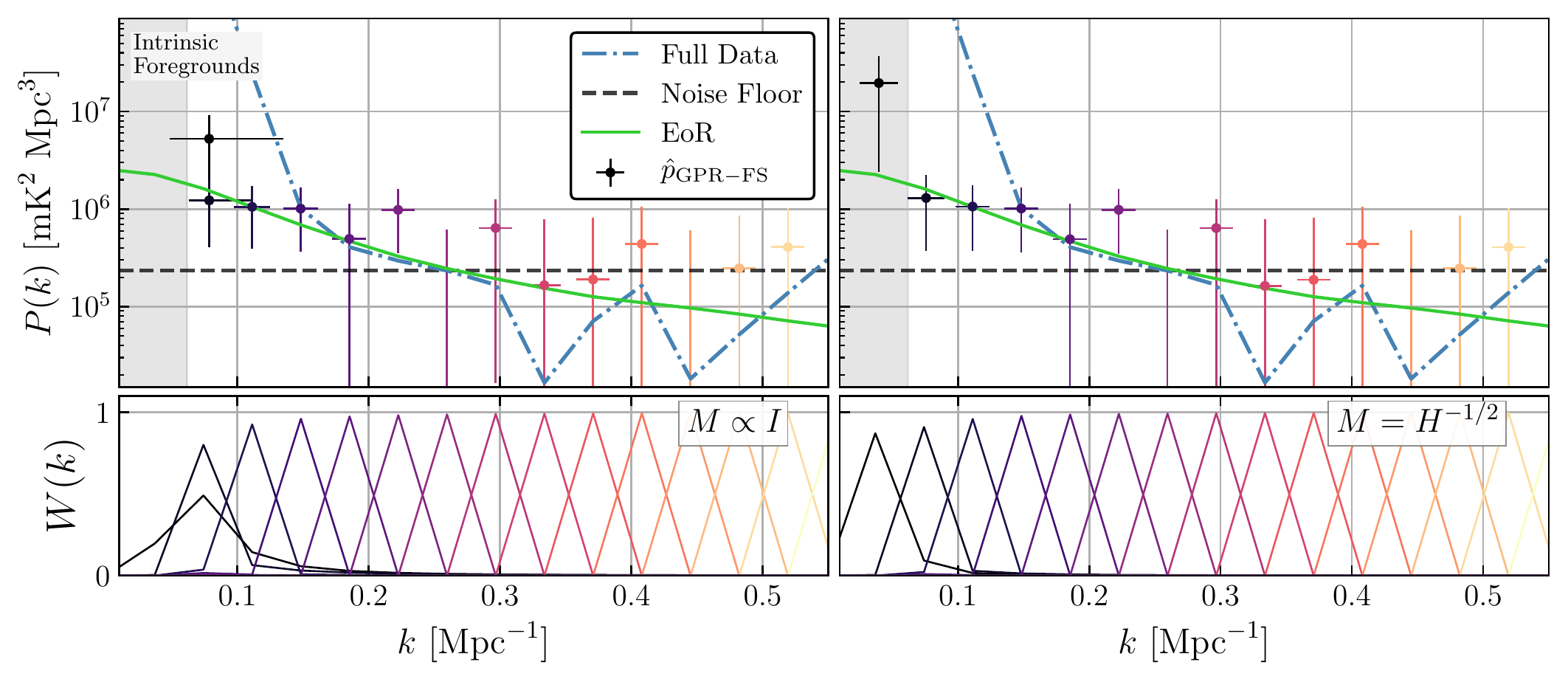}
\caption{The same as \autoref{fig:gpr_window_med_noise} but with a foreground term that is weaker by an overall factor of $10^4$ in power (and thus occupies a smaller range in $|k|$ than before).
While this is somewhat unrealistic for a real experiment, this could be in practice be achieved via a direct foreground subtraction step before applying GPR-FS, although understanding the residual covariance of the foregrounds is itself a non-trivial endeavor. Not surpisingly, GPR-FS is able to probe the EoR power spectrum to below $k < 0.1\ h\ {\rm Mpc}^{-1}$ with considerably tighter errorbars in such a scenario.}
\label{fig:gpr_window_med_noise_weakfg}
\end{figure*}

\subsection{Mapping the Window Functions}
\label{sec:gpr_windows}

One of the blessings of incorporating GPR modeling into the quadratic estimator formalism is the ability to propagate its effect on the band power covariance and window function matrices.
Recall that the window functions tell us how the intrinsic band powers map into our measured band powers.
In a sense, they represent the true $k$ modes that our measured band powers are sensitive to and the width of the band power bin (i.e. the horizontal error bar).

To demonstrate an idealized case of GPR-FS and to map its window functions, we generate mock visibility data from the foreground, EoR, and noise covariances described in \autoref{sec:gpr}.
\autoref{fig:data_covariance} shows the covariances of our three data terms.
We simulate the visibility for a single baseline spanning 20 MHz of bandwidth anchored at 155 MHz in 64 frequency channels by drawing from a mean-zero complex Gaussian distribution with the covariances described above for each of the foreground, EoR, and noise terms, before summing them together to form our data vector $\d$.
For power spectrum estimation, we form two quantities, $\d_1 = \f + \e + \n_1$ and $\d_2 = \f + \e + \n_2$, which have the \emph{same} foreground and EoR realization but independent noise realizations.
These terms enter the left-hand and right-hand side of \autoref{eq:qe} such that our final power spectra are free of the noise-bias component, a common analysis technique when working with real data.
We repeat this 1000 times (drawing an independent realization of all terms each time) and stack them to construct a data vector $\d$ of shape $(N_{\rm freqs}, N_{\rm realizations}) = (64, 1000)$.
Because this uses the same covariance for each draw, one can think of this as drawing over a combination of different baseline orientations (of similar length) and patches of the sky, which will help average down sample variance of the signal terms in the estimated power spectrum.
We use the \texttt{simpleQE} package\footnote{\url{https://github.com/nkern/simpleQE}} for power spectrum estimation, which is a simple, straight-forward implementation of the equations in \autoref{sec:qe}.

We simulate two scenarios that will be relevant for future 21\,cm experiments: one where the EoR signal is detected at high significance with a signal-to-noise ratio (SNR) > 10 (low-noise scenario) and another where the EoR is marginally detected with SNR = 1 (moderate noise scenario).
The low-noise scenario is shown in \autoref{fig:gpr_window_low_noise}, where we show the recovered power spectrum (points) after GPR-FS with $\M\propto\I$ normalization (left) and $\M=\H^{-1/2}$ normalization (right), as well as their window functions (bottom).
We also show the full dataset including foregrounds, EoR, and noise (blue dot-dashed), the intrinsic EoR component (green) and the noise floor (black dashed).
At high $k$ where the thermal noise dominates the two schemes give nearly the same answer; however, at low $k$ the two normalization schemes yield substantially different results.
While $\M\propto\I$ normalization leads to smaller vertical errorbars, we see that it is not actually that sensitive to the power spectrum at $k \lesssim 0.13\ h\ {\rm Mpc}^{-1}$.
While the $\M=\H^{-1/2}$ normalization scheme has larger vertical errorbars at low $k$, it is in principle probing the power spectrum at those $k$ modes.
If we had not used the window function to inform the horizontal placement of the points in \autoref{fig:gpr_window_low_noise}, we would have been erroneously led to believe we are actually probing the power spectrum at $k \le 0.1\ h\ {\rm Mpc}^{-1}$ with the $\M\propto\I$ normalization.
Perhaps not surprisingly, similar effects were seen with the inverse covariance quadratic estimator in \citet{Liu2014b}.

\autoref{fig:gpr_window_med_noise} shows the same test but in a moderate-noise scenario, which is the more realistic case for a first-detection of the 21\,cm power spectrum by an experiment like HERA or the SKA.
We see roughly the same trends as before between the two normalization schemes.
GPR-FS with $\M\propto\I$ normalization is unable to actually probe the power spectrum at very low $k$, although its window functions are marginally better behaved at these modes, while a more careful deconvolution of the window function leads to band powers that are able to probe low $k$ modes at the expense of larger vertical errorbars.

The major reason why GPR-FS is failing to accurately recover the EoR at low $k$ in both the low and moderate noise scenario is simply due to the large dynamic range between the foregrounds and the EoR at those Fourier modes.
We can show this by repeating the moderate-noise GPR-FS test with an artifically weakened foreground signal that is $10^4$ times weaker in power.
While perhaps not realistic for any actual experiment, this could in practice be achieved by applying a direct foreground subtraction step before GPR-FS, although modeling the residual foreground covariance after such a step would also be a fairly non-trivial endeavor.
Nonetheless, \autoref{fig:gpr_window_med_noise_weakfg} shows the result of GPR-FS on this dataset, demonstrating improved performance in its ability to recover the EoR signal at $k\sim0.1\ h\ {\rm Mpc}^{-1}$ for both normalization schemes.
In this scenario, the window functions are much better behaved at these $k$ modes (although $\M\propto\I$ still cannot probe very low $k$ modes), with the added bonus of significantly tighter vertical errorbars.

\begin{figure*}
\centering
\includegraphics[width=\linewidth]{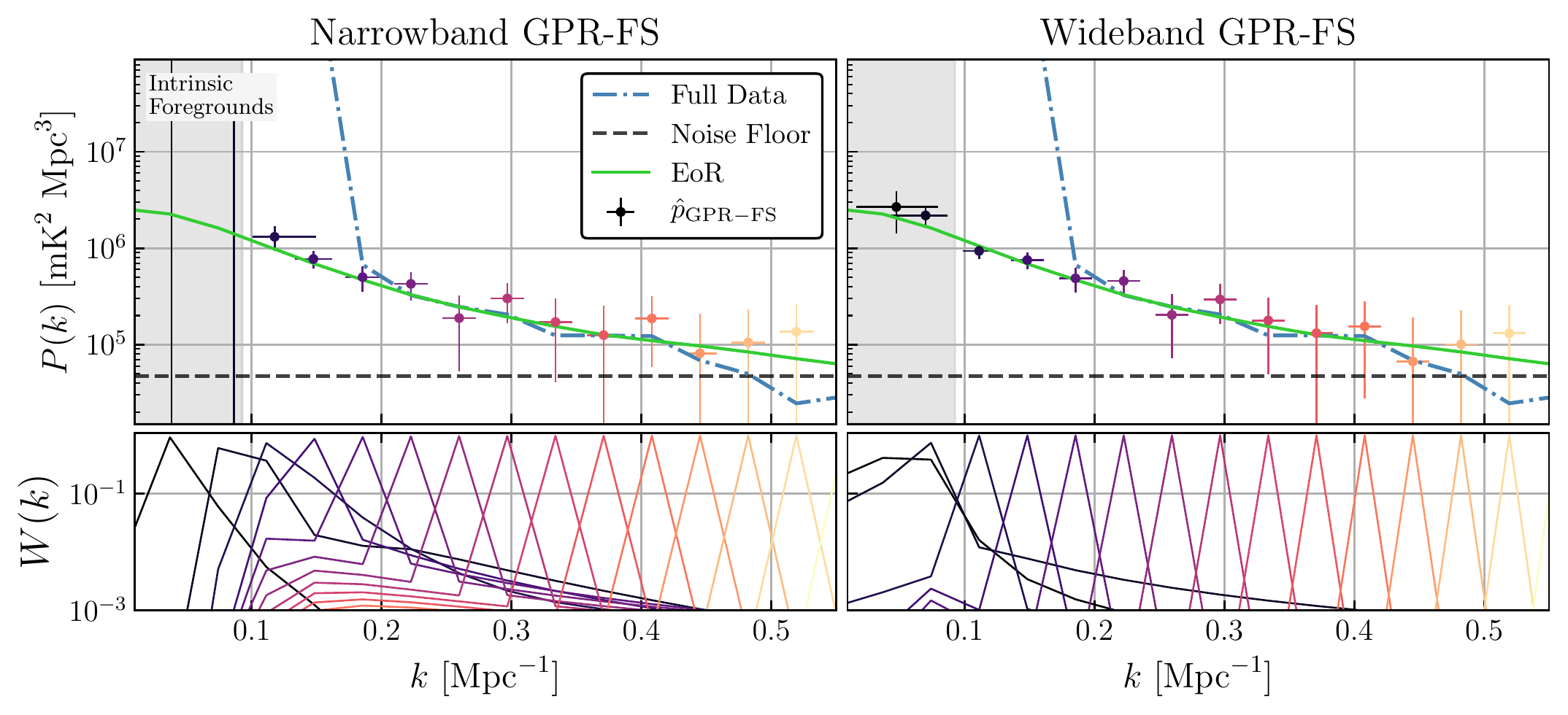}
\caption{GPR-FS applied to a 140 -- 160 MHz bandwidth (left) and wideband GPR-FS applied to a 110 -- 190 MHz bandwidth but with a power spectrum spectrum window of 140 -- 160 MHz (right), both with $\M=\H^{-1/2}$ normalization.
We plot the window functions on a log scale here to highlight the low-level differences between narrowband and wideband GPR-FS.
Compared to the narrowband case, wideband GPR-FS exhibits sharper and less correlated window functions for Fourier modes down to $k\sim0.1\ h\ {\rm Mpc}^{-1}$.
For Fourier modes $k \le 0.1\ h\ {\rm Mpc}^{-1}$, wideband GPR-FS exhibits considerably tighter vertical errorbars but is unable to significantly shrink the horizontal extend of the window functions.}
\label{fig:gpr_wideband}
\end{figure*}

\section{Extensions to GPR-FS}
\label{sec:extensions}

In this section, we describe two simple extensions to the GPR-FS technique described above.
These extensions improve the performance of GPR-FS and help to address the real-world issue of missing or nulled data, which can be neatly folded into the quadratic estimator formalism.

\subsection{Wideband GPR-FS}
\label{sec:wideband_gpr}

Here we explore the ability of GPR-FS to model and downweight foregrounds across a wide bandwidth, while simultaneously estimating the power spectrum across the same narrow bandwidth from before.
Modeling foregrounds across a wide bandwidth is desireable because it increases the resolution in Fourier space with which the foregrounds can be removed.
This has proven beneficial in practical data analyses of foreground filtering and modeling \citep{Parsons2014, Ali2015, Carucci2020}.
This also has the benefit of decreasing the correlation between neighboring Fourier modes in the recovered power spectrum.
\citet{Ewall-Wice2020}, for example, demonstrates how wideband foreground filtering, as opposed to narrowband filtering, can improve the indepedence of the window functions in Fourier space.

There are a few complications, however, for a wideband GPR-FS.
The first is the increased computational complexity, as GPR naively scales as $N_\nu^3$.
For most 21\,cm experiments, however, a wideband GPR-FS spanning $\sim100$ MHz will operate over $\ge1024$ channels, which should not be prohibitively expensive, especially given standard GPR optimizations that allow it to scale roughly as $\mathcal{O}(N^2)$ for medium-sized datasets \citep{Rasmussen2006}.
The second is the fact that many of the assumptions made previously about the stationarity of the signal begin to break down over large bandwidths, particularly in the amplitude of the foregrounds and thermal noise, which increase at low frequencies.
A simple power-law extension to the covariances discussed above should be a straightforward way to deal with this issue, which we defer to future work.
Lastly, the 21\,cm signal itself evolves along the line-of-sight probed by the frequency axis, which is why power spectrum analyses generally limit themselves to bandwidths $<10$ MHz.

We can, however, apply a wideband GPR-FS and simultaneously estimate the 21\,cm power spectrum across a narrow bandwidth while keeping everything contained within the QE framework, thus preserving the nice statistical properties of the bandpower covariance and the window functions.
Similar to \citet{Ewall-Wice2020}, we start by forming a large GPR-FS filtering matrix the size of the wider bandwidth.
We then truncate $\R_{\rm GPR-FS}$ on one side, yielding a non-square weighting matrix, with a dimensionality of $N_{\rm spw}\times N_{\nu}$, where $N_{\rm spw} < N_\nu$ and $N_{\rm spw}$ is the number of channels in our narrowband spectral window.

We run wideband GPR-FS in the same manner as before, but simulate a wider bandwith spanning 110 -- 190 MHz with 256 frequency channels, four times the bandwidth as before.
However, we still use the same 20 MHz spectral window for estimating the power spectrum and normalize the power spectra with $\M=\H^{-1/2}$.
The results are found in \autoref{fig:gpr_wideband}, which shows both a decrease in the width of the window functions and their correlation floor when moving to wideband GPR-FS.
For the low $k$ band powers their vertical errorbars are significantly reduced, however, their window functions still span a fairly broad region in $k$.

\begin{figure*}
\centering
\includegraphics[width=\linewidth]{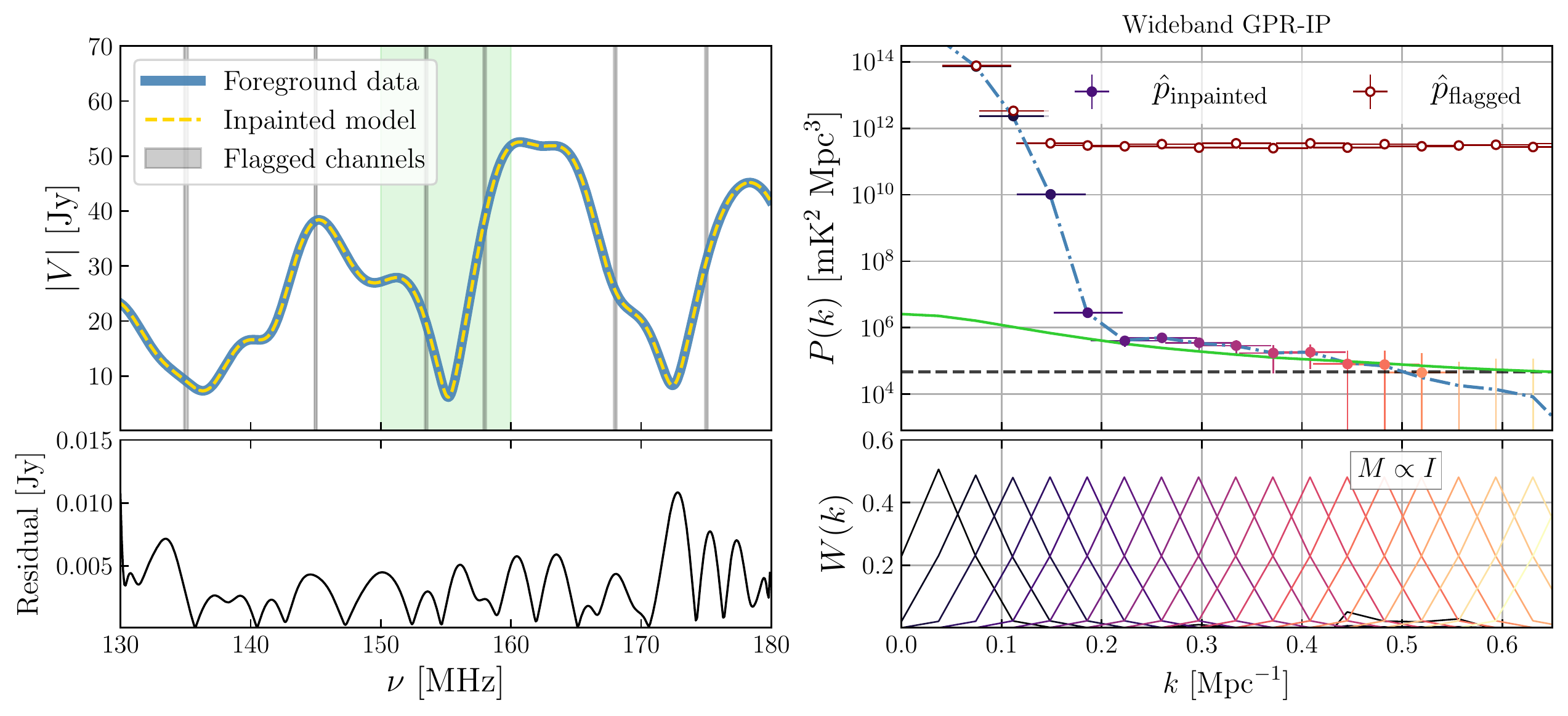}
\caption{GPR inpainting applied to the mock dataset from above. Although we only show the foreground data, \autoref{eq:gprip} is applied to the full dataset to construct the foreground model. {\bfseries Left}: We show the initial foreground data (blue) with flags applied (shaded) and the inpaint model (yellow dashed). The residual between the inpaint foreground model and the initial foreground data is also plotted (bottom). While the entire band is used for inpainting, the green shaded region indicates the power spectrum subband. {\bfseries Right}: We plot the inpainted power spectrum (points) compared to the original data (blue), EoR signal (green) and the noise floor (black dashed), along with the non-inpainted band powers (red). All power spectra are normalized with a $\M\propto\I$ convention.}
\label{fig:gpr_inpaint}
\end{figure*}

\subsection{Inpainting Missing Data}
\label{sec:inpainting}

Often when working with real instruments we must excise or flag data due to poor quality.
This can motivated by detector or instrument failures or by contamination of certain parts of the data, for example due to radio frequency interference.
Intensity mapped power spectra are particularly sensitive to missing data along the frequency axis, as the Fourier transform of discontinuous features will cause ringing of bright Fourier modes (such as foregrounds) to other neighboring modes, thus contaminating them.
In theory, decorrelating the band powers with the normalization step in the quadratic estimator (i.e. setting $\M=\H^{-1}$) can help with this, but the impact of missing data can often lead to ill-conditioned matrices in the quadratic estimator, as the number of estimated parameters (the band powers) exceeds the number of observations.

Another approach for handling the missing data problem is to fill-in the nulled data with our best guess of the signal, called data inpainting.
Inpainting can help prevent these features from contaminating other modes in the first place.
In practice we are less concerned about the EoR signal ringing against itself in Fourier space, as most theoretical models predict fairly flat features in $P(k)$, but in general this is still a concern, and eliminating the ringing of the EoR signal can help to prevent unnecessary correlation between nearby band powers of high SNR detections.
Another reason for wanting to inpaint the data is because we often want to make power spectra of the foreground component, as this is a useful metric for assessing data quality and performing jacknife and null tests.

An example of a well-known radio astronomy technique for solving this problem is the Hogbom CLEAN algorithm and its derivatives \citep{Hogbom1974}.
A 1D variation of this algorithm applied along the frequency axis of radio visibilities has been extensively used in PAPER and HERA analyses \citep{Parsons2009, Ali2015, Kerrigan2018, Kern2020a}.
\citet{Offringa2019} also demonstrated an inpainting technique for dealing with RFI in LOFAR data, and \citet{Ewall-Wice2020} recently presented a visibility filter that uses a delay-space maximum likelihood method for inpainting missing data.
Inpainting also has a long history in CMB analyses \citep[e.g.][]{Starck2013, Gruetjen2017}.
Recently, \citet{Trott2020} explored how GPR inpainting (aka `kriging') can be used to partially mitigate the effect of regularly contaminated frequency bins in MWA data, however, their formulation was somewhat ad-hoc and did not seek to model the foreground covariance directly, and possibly as a consequence exhibited biases at high $k$ modes.
Here, we demonstrate how data inpainting can be performed using much of the same GPR framework from before while seamlessly incorporating it into the quadratic estimator, thereby including its effect into the full statistical description of the band power window function and covariance matrix.

Recall \autoref{eq:gp_conditional} gives the distribution of the desired latent variable conditioned on the data.
For the purposes of foreground subtraction, the desired latent variable was assumed to be the foreground component sampled at the same frequencies as the data (\autoref{eq:fg_conditional}).
For the purpose of inpainting we can relax some of these assumptions.
First, inpainting should only impact the data voxels with missing data, and should not alter otherwise cleanly sampled data.
Therefore, the output of the GP model should be summed with the data vector only at the flagged frequency bins.
Furthermore, we do not want the GP conditioned on the missing data, as it is not part of the statistical distribution we are trying to model.
We can handle this by either giving those frequency bins an infinite noise variance (thus in practice eliminating their contribution to the data model), or we can simply truncate them from the data vector $\d$, as the GP need not sample the data at regularly spaced intervals.
To be consistent with $\nu$ representing the native data frequencies and the input frequencies to the GP, we will adopt the former convention.
Second, we may want to inpaint multiple components of the data, not just the foreground signal.
To capture this, we define $\dip$ as the latent variable to-be-inpainted in frequency space, and $\Kip$ as our model for its covariance.

This leads us to construct a generic Gaussian process data model matrix (GPR-DM)
\begin{align}
\label{eq:gprdm}
\R_{\operatorname{GPR-DM}} = \Kip[\Kfg + \Kto + \Kn + \sigma_{\rm f}^2\W_{\rm f}]^{-1},
\end{align}
where $\W_{\rm f}$ is a diagonal matrix that is zero at unflagged channels and one at flagged channels, and $\sigma_{\rm f}^2$ is a pre-determined large number.
This gives us the GPR estimate at all data frequencies.
Note that the act of $\sigma_{\rm f}^2\W_{\rm f}$ is to add exceedingly large variance to the specific channels that are flagged, so long as $\sigma_{\rm f}^2$ is considerably larger than the other terms in the full data covariance, such that those channels do not influence the overall GPR fit.
We can combine this with the appropriate weighting matrices to construct an inpainting matrix,
\begin{align}
\label{eq:gprip}
\R_{\operatorname{GPR-IP}} = \I - \W_{\rm f} + \W_{\rm f}\R_{\operatorname{GPR-DM}}.
\end{align}
Note that this special $\R$ matrix dotted into our data vector does not affect the channels that are not flagged in the data: it merely fills-in the flagged channels with the GPR fit.

When forming power spectra of data with a large dynamic range between Fourier modes, it is often beneficial to apply a tapering function that smoothly connects that data to zero at the band edges (also known as apodization), which limits the Fourier space ringing induced by the fact that our data are non-periodic over our finitely sampled bandwidth.
We can do this adopting $\T\R_{\operatorname{GPR-IP}}$ as our weighting matrix, where $\T$ is a diagonal matrix with a tapering function applied to its diagonal.
A commonly adopted tapering function, which we adopt here, is the Blackman-Harris function that suppresses Fourier sidelobe structure by five orders of magnitude \citep{Blackman1958}.

In \autoref{fig:gpr_inpaint}, we show an example of this estimator applied to the same data products discussed above, where we are attempting to inpaint the foreground signal and thus measure the power spectrum of both the foreground and EoR signal.
The left panel shows the foreground model constructed from $\R_{\operatorname{GPR-DM}}$ (yellow dashed) compared to the initial foreground data (blue), along with the flagged regions (grey shaded).
The right panels shows the estimated power spectrum with and without inpainting and the inpainted band power window functions.
Inpainting helps us recover accurate estimates of the foreground power over spectral windows with missing data, which is useful for data quality management and null testing of data analysis pipelines \citep{Kolopanis2019}.
Note that the fidelity of the reconstruction will be dependent on the bandwidth and the flagging occupancy of the data: more flagged data leads to poorer reconstruction. A detailed study of this dependence is outside the scope of this work, which we leave to future instrument and RFI environment specific studies.

A key distinction between the GPR inpainting technique described here and the GPR-FS algorithm described previously is found by inspecting their respective window functions.
\autoref{fig:gpr_inpaint} shows significantly better-behaved window functions at low $k$ modes for $\M\propto\I$ normalization when compared to GPR-FS (e.g. \autoref{fig:gpr_window_med_noise}).
This difference is driven by the fact that in our inpainting scenario, we are attempting to estimate the joint foreground and EoR power spectrum, which doesn't require high dynamic range signal separation at low $k$.
This is a reminder that the resultant window functions of our band powers are not merely driven by our choice of power spectrum estimator and normalization scheme, but also in our choice of the signal we aim to estimate.


\section{LOFAR 2020 Limits As a Case Study}
\label{sec:lofar}
In this section, we look at the implementation of GPR foreground removal in a recent LOFAR analysis of the 21\,cm power spectrum \citepalias{Mertens2020}, and in particular seek to understand how GPR modeling is affecting their recovered power spectra at the lowest $k$ modes.
This has particular importance given recent astrophysical parameter constraints derived from these limits \citep{Ghara2020, Mondal2020, Greig2020}, which are driven primarily by the low $k$ modes that have the largest signal-to-noise.
In summary, when casting the normalization scheme of \citetalias{Mertens2020} into the QE framework, we find that they make a tradeoff between normalization and bias correction, which lends their estimator to being particularly sensitive to EoR signal loss\footnote{We define EoR signal loss as any analysis step that leads to the underestimation of the EoR power spectrum, whether this be due to calibration, filtering, or improper power spectrum normalization.} when their adopted EoR model is misestimated.
To demonstrate this, we look at how well the EoR power spectrum is recovered when the assumption of a known data covariance is relaxed, and is instead regressed for over the space of the covariance hyperparameters.
In the limit that the true EoR covariance cannot be described by the EoR covariance model, we find that the \citetalias{Mertens2020} normalization scheme can suffer from significant signal loss at low $k$ modes.
While we do not look at them specifically, other analyses utilizing the LOFAR GPR foreground removal pipeline may suffer from similar issues discussed in this section \citep[e.g.][]{Gehlot2018, Ghosh2020}.

\subsection{Power Spectrum Recovery Tests}
\label{sec:bias_correction}


In \citetalias{Mertens2020} Sec. 3.3.2 it is argued that the inferred variance of the residual vector after GPR foreground subtraction is a biased estimate (biased low) of the noise and 21\,cm components, and can be corrected by summing the inferred variance with the analytic variance on the GP foreground estimate (\autoref{eq:fg_cov_conditional}).
However, this result hinges on a crucial assumption that the adopted data covariance is in fact equal to the true covariance of the data.
To understand the impact this has on the resultant power spectra, we can frame it in terms of the standard QE normalization and bias subtraction steps.
In \autoref{sec:bias_correction_math}, we outline the differences in the normalization and bias subtraction steps between the standard QE and the estimator from \citetalias{Mertens2020}.
We show that \citetalias{Mertens2020}'s estimator is particularly sensitive to errors in the adopted 21\,cm covariance, which is a result of the additional assumption made in their normalization step that the data covariance is matched to the true covariance.
 
\begin{figure}
\centering
\includegraphics[width=\linewidth]{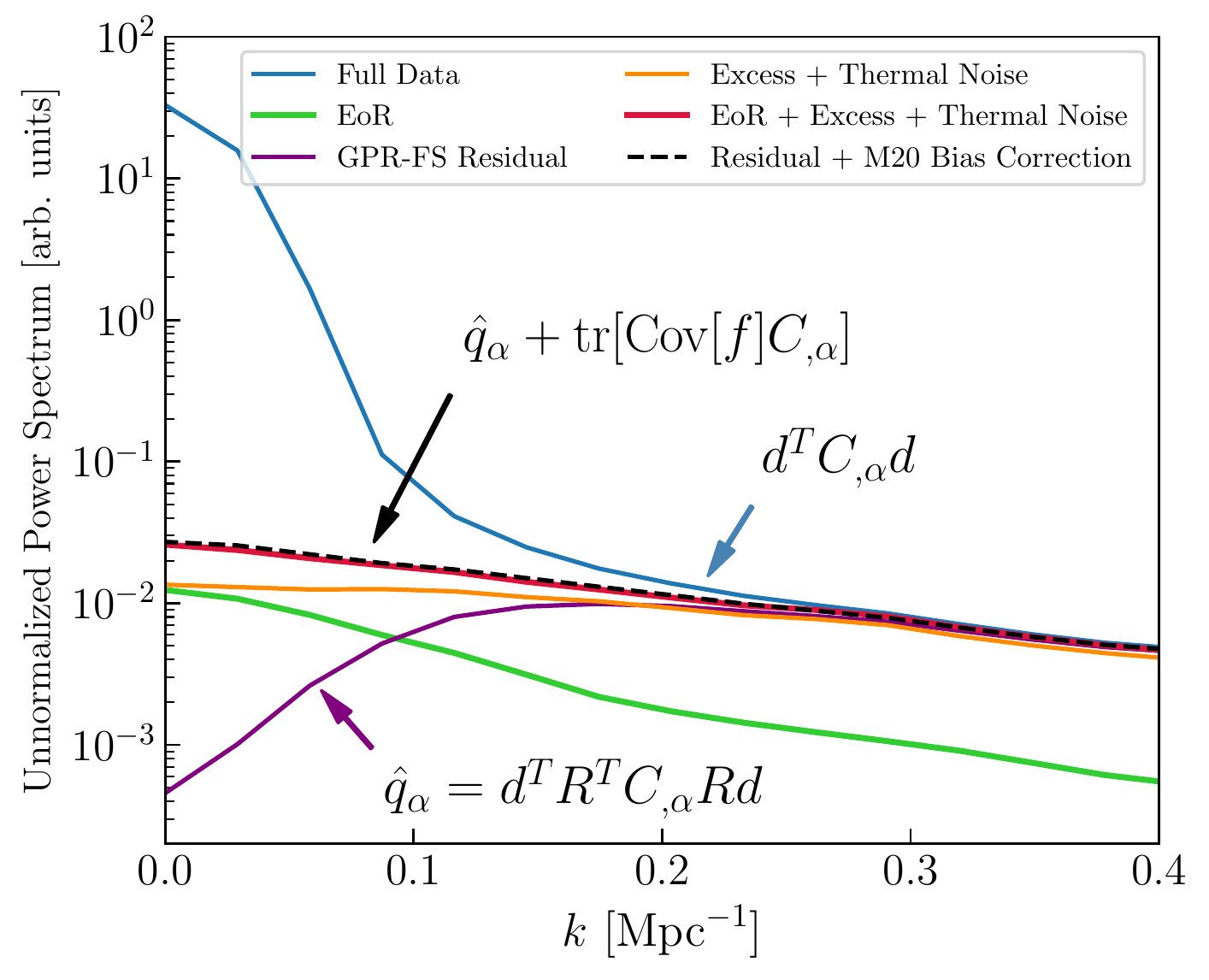}
\caption{The unnormalized power spectrum of a realization of the LOFAR dataset (blue), its GPR-FS residual (purple), and the residual summed with the bias correction term (dashed black).
We also show the intrinsic EoR component (green), the excess and thermal noise component (orange) and their sum (red).
We see that the GPR-FS operator filters off not only foregrounds but also noise and EoR at low $k$ modes.
When we know the true covariance models of the data (as we do here, for test purposes) then $\hat{q}_\alpha + \tr[\Cov[\f]\C_{,\alpha}]$ does indeed form an unbiased estimator of the EoR and noise components (demonstrated by the agreement between the red and black dashed) as \autoref{eq:res_covariance} would suggest.}
\label{fig:lofar_resid}
\end{figure}

We can further demonstrate this via simulated power spectrum recovery tests.
To do this, we generate mock realizations of the LOFAR dataset from \citetalias{Mertens2020} by drawing a series of Gaussian random fields from each of the covariances outlined in their Table 3, before summing them together.
This includes a foreground component, an EoR component, a thermal noise component, and an excess noise component.\footnote{\citetalias{Mertens2020} find evidence for an ``excess noise'' component of their data, whose origin they cannot fully explain, although they speculate that it could be caused by antenna-to-antenna mutual coupling.}
We repeat this 200 times to simulate independent visibilities (with the same covariance parameters), which are summed together after forming their power spectra.
A notable exception: in order to actually test our ability to recover the EoR signal, we boost the EoR covariance amplitude over what is reported in their Table 3 such that it exceeds the excess noise component at low $k$.

Our first test is a null test of sorts, and asks whether we can recover an unbiased estimate in the limit that we know the true data covariance.
In \autoref{fig:lofar_resid}, we show the unnormalized power spectrum of the full data (blue), the residual after applying GPR-FS (purple), the true EoR component (green), and the power spectrum estimate using the bias correction method of \citetalias{Mertens2020} (black dashed).
Importantly, \autoref{fig:lofar_resid} shows that GPR-FS filtering does indeed inherently subtract non-foreground terms, particularly at low $k$ where we see it subtracting off power from the intrinsic EoR and noise components.
In the case where we know the true covariances, however, the residual plus the bias correction (black dashed) does make an unbiased estimate of the EoR and noise components across all $k$.

Next we ask what happens when we relax the assumption of a known covariance, and actually regress for the covariance hyperparameters by maximizing the marginal likelihood and prior (holding the noise fixed) with a gradient descent algorithm.
All hyperparameter priors are flat with hard bounds given by \citetalias{Mertens2020}'s Table 3.
Of particular note, \citetalias{Mertens2020} adopt a prior on the 21\,cm covariance length scale $\ell_{21}$ between $[0.1, 1.2]$ MHz, which they claim does not impact the maximum marginal likelihood (private communication).

\begin{figure*}
\centering
\includegraphics[width=\linewidth]{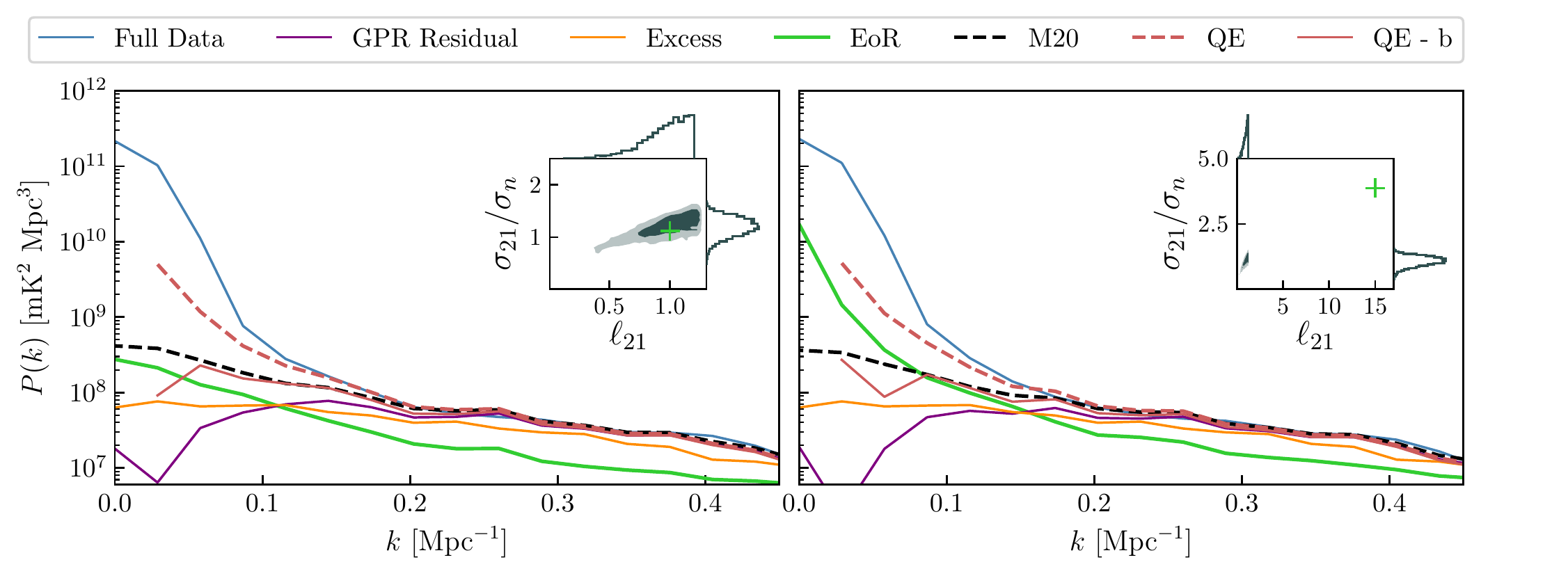}
\caption{
Power spectra of EoR recovery tests on mock LOFAR datasets while changing the EoR covariance model length scale parameter.
We show the full dataset (blue), the GPR residual (purple), the excess noise (orange), the EoR component (green), the \citetalias{Mertens2020} normalized spectra (black dashed), the QE normalized spectra without foreground bias subtraction (red dashed), and the QE estimate with foreground bias subtraction (red solid).
The QE power spectra are normalized with an $\M=\H^{-1/2}$ scheme, and are truncated below $k < 0.03\ {\rm Mpc}^{-1}$ because the window functions are not locally compact in $k$.
The thermal noise bias is subtracted off in all cases, but the excess noise bias is not.
All covariance hyperparameters (except for the thermal noise amplitude) are re-optimized in each case, with inset showing the marginalized 68\% and 95\% confidence intervals of the marginal likelihood across the two EoR covariance hyperparameters.
We see that when the EoR covariance falls within the scope of the model, both the \citetalias{Mertens2020} and the QE estimator (without bias subtraction) report power spectra consistent with a claimed upper limit on the signal; however, when the true EoR covariance is taken outside the bounds of the model, the \citetalias{Mertens2020} normalization suffers from signal loss for $k < 0.08\ {\rm Mpc}^{-1}$.
We also see that the QE estimate with bias subtraction (red solid) suffers from some signal loss due to imperfectly estimated foreground covariance parameters.
}
\label{fig:lofar_pspec_prior_test}
\end{figure*}

\begin{figure*}
\centering
\includegraphics[width=\linewidth]{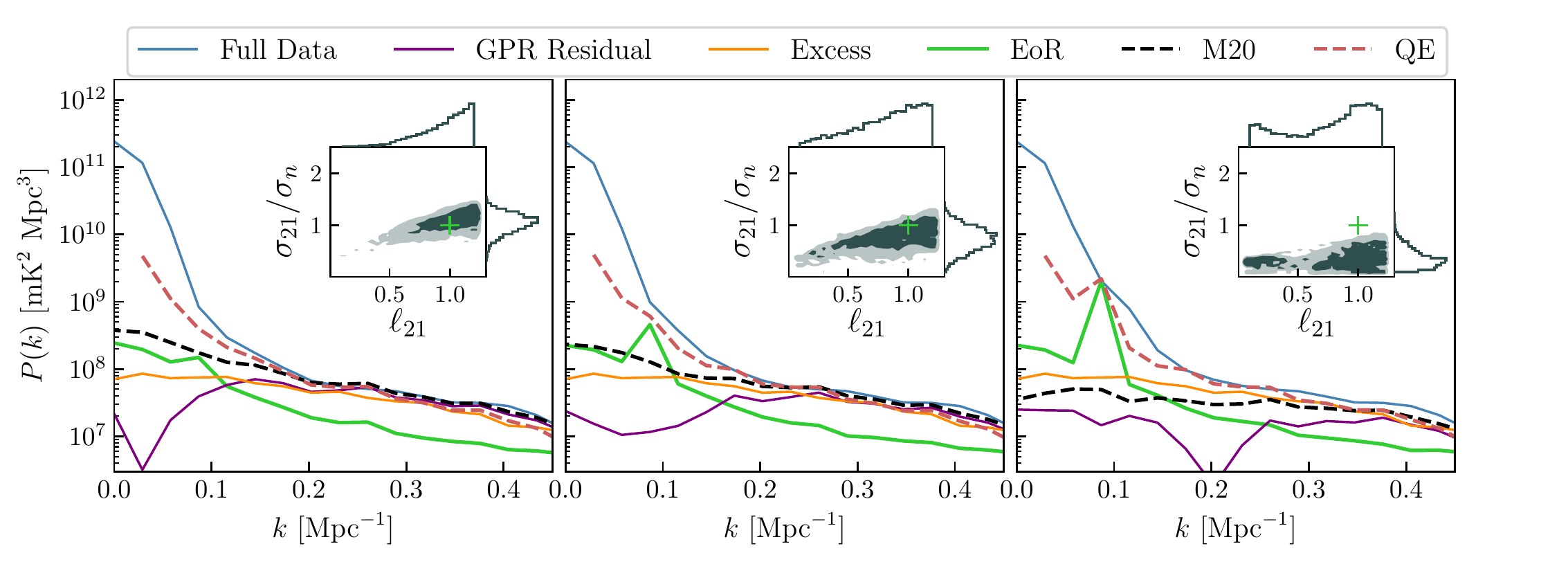}
\caption{Recovered power spectra when inserting a tone into the EoR signal at $k=0.085\ {\rm Mpc}^{-1}$ of increasing amplitude (left to right). While the \citetalias{Mertens2020} estimator is unable to respond to the tone, the QE estimator does respond to it, and is able to set robust upper limits across all the $k$ modes probed. While a tone structure is somewhat contrived, many theoretical EoR models do have a rolloff knee feature low $k$ \citep{Greig2015a}, which may pose similar problems to the one seen here. Here we do not plot the QE estimate with the foreground bias subtraction as we expect the foreground covariance to be misestimated due to the additional variance, although similar to \autoref{fig:lofar_pspec_prior_test} it roughly traces the \citetalias{Mertens2020} estimator.}
\label{fig:lofar_pspec_tone_injection}
\end{figure*}

We simulate two scenarios here, one with $\ell_{21}=1$ MHz falling within the prior bounds, and another with $\ell_{21}=15$ MHz falling outside the prior bounds (\autoref{fig:lofar_pspec_prior_test}).
We show the recovered power spectra with \citetalias{Mertens2020}'s normalization (black dashed), with the QE normalization without foreground bias subtraction (red dashed), and the QE estimate with bias subtraction (red solid).Furthermore, we truncate the QE estimate below $k < 0.03\ {\rm Mpc}^{-1}$ because its window functions are not locally compact.
The thermal noise bias is subtracted off in both cases, but the excess noise bias is not, as is consistent with \citetalias{Mertens2020}.

The left panel of \autoref{fig:lofar_pspec_prior_test} shows the recovered power spectra in the limit that the true $\ell_{21}$ lies within the parameter bounds (green cross), shown in the inset.
The shaded contours represent the 68 and 95\% confidence interval of the hyperparameter posterior distribution\footnote{Although we only show the EoR covariance parameters here for simplicity, all covariance hyperparameters are regressed for jointly.} showing it running up against the hard prior set at 1.2 MHz, similar to that observed in Figure B1 of \citetalias{Mertens2020}.
In this case, both the \citetalias{Mertens2020} and non-bias-subtracted QE estimator are free of EoR signal loss, however, the latter estimator also has significant excess due to the lack of a foreground bias subtraction.
Both would be correct in claiming upper limits on the EoR signal.
The foreground bias subtracted QE estimate (red solid) does suffer some signal loss, due to imperfectly estimated foreground parameters.

When we move the true EoR covariance parameters outside of the prior bounds (right panel), we see a different response.
The covariance hyperparameters are re-regressed in this scenario, but because of the hard prior at 1.2 MHz the EoR model is mismatched from the true EoR model, which leads to significant underprediction of the EoR power for $k < 0.8\ {\rm Mpc}^{-1}$ in the \citetalias{Mertens2020} estimator.
At the lowest $k$ bin quoted by LOFAR of $k=0.5\ {\rm Mpc}^{-1}$, this discrepancy is upwards of a factor of two.
The QE estimate (without bias subtraction) on the other hand, is still correct in claiming an upper limit on the signal, even having adopted a mismatched EoR covariance.

Another way to test how a mismatched covariance impacts the recovered power spectra is to insert a tone into the EoR signal (without changing the functional form of the EoR covariance) and seeing out the recovered power spectra respond.
In \autoref{fig:lofar_pspec_tone_injection} we show this where the base EoR model parameters lie within our parameter space (green cross), but now we insert a tone into the EoR data at $k=0.085\ {\rm Mpc}^{-1}$ of increasing variance from left to right, showing how the fitted EoR covariance parameters respond, and how the recovered power spectra respond.
We see a systematic shift in the fitted EoR hyperparameters due to a degeneracy between the tone amplitude and the EoR hyperparameter space, but more importantly we see that the \citetalias{Mertens2020} power spectra are largely insensitive to the tone.
The QE estimate though, even with a severely mismatched EoR covariance, does respond to the tone, and is able to set robust upper limits on the signal across all $k$ modes probed.
While a tone is somewhat contrived, many EoR models exhibit a rolloff knee feature at low $k$ \citep{Greig2015a}, which may create similar problems for estimators that have trouble passing the test demonstrated in \autoref{fig:lofar_pspec_tone_injection}.

It is important to note that had we subtracted the bias term of the QE in \autoref{fig:lofar_pspec_prior_test} and \autoref{fig:lofar_pspec_tone_injection}, it would have suffered from similar issues seen with the \citetalias{Mertens2020} estimator.
Our conclusions from this section are simply that, in the conservative case where we do not completely trust our data covariance models, the QE estimator can set robust upper limits on the signal (while still suppressing some foreground power) even with fairly mismatched covariances.
With the bias correction employed by \citetalias{Mertens2020}, however, we find that it can suffer from signal loss when the EoR covariance is mismatched from the true covariance, whether it is because we adopted a hard prior the limits the parameter space (\autoref{fig:lofar_pspec_prior_test}), or because our model simply cannot capture the full degrees of freedom of the EoR data (\autoref{fig:lofar_pspec_tone_injection}).

In fairness, \citetalias{Mertens2018} and \citetalias{Mertens2020} employ a wide range of tests, both with simulated and real data, to show that their estimator is robust against signal loss, and they clearly acknowledge that getting a correct data covariance model is key to their analysis.
To this end, \citetalias{Mertens2020} explore a few different covariance models for the excess noise and foreground components, picking the one that maximized the Bayesian evidence (their Table 2), and also claim that the prior on $\ell_{21}$ of [0.1, 1.2] MHz did not dramatically change their hyperparamter posterior distribution (private communication).
However, the key distinction made by this work is that the covariance model selection only works when the true EoR covariance falls within a subspace of the adopted EoR covariance model, regardless of whether we allow the hyperparameters of that model to find a local minimum.
Furthermore, in the simulated recovery tests of \citetalias{Mertens2020}, we argue that they begin to see hints of this effect in their Figure A2, which shows increasing signal loss for $k<0.15\ {\rm Mpc}^{-1}$ as their input EoR model length scale $\ell_{21}\rightarrow 1.2$ MHz.

\subsection{Implications for Astrophysical Constraints}
\label{sec:astro_params}

\begin{figure}
\centering
\includegraphics[width=\linewidth]{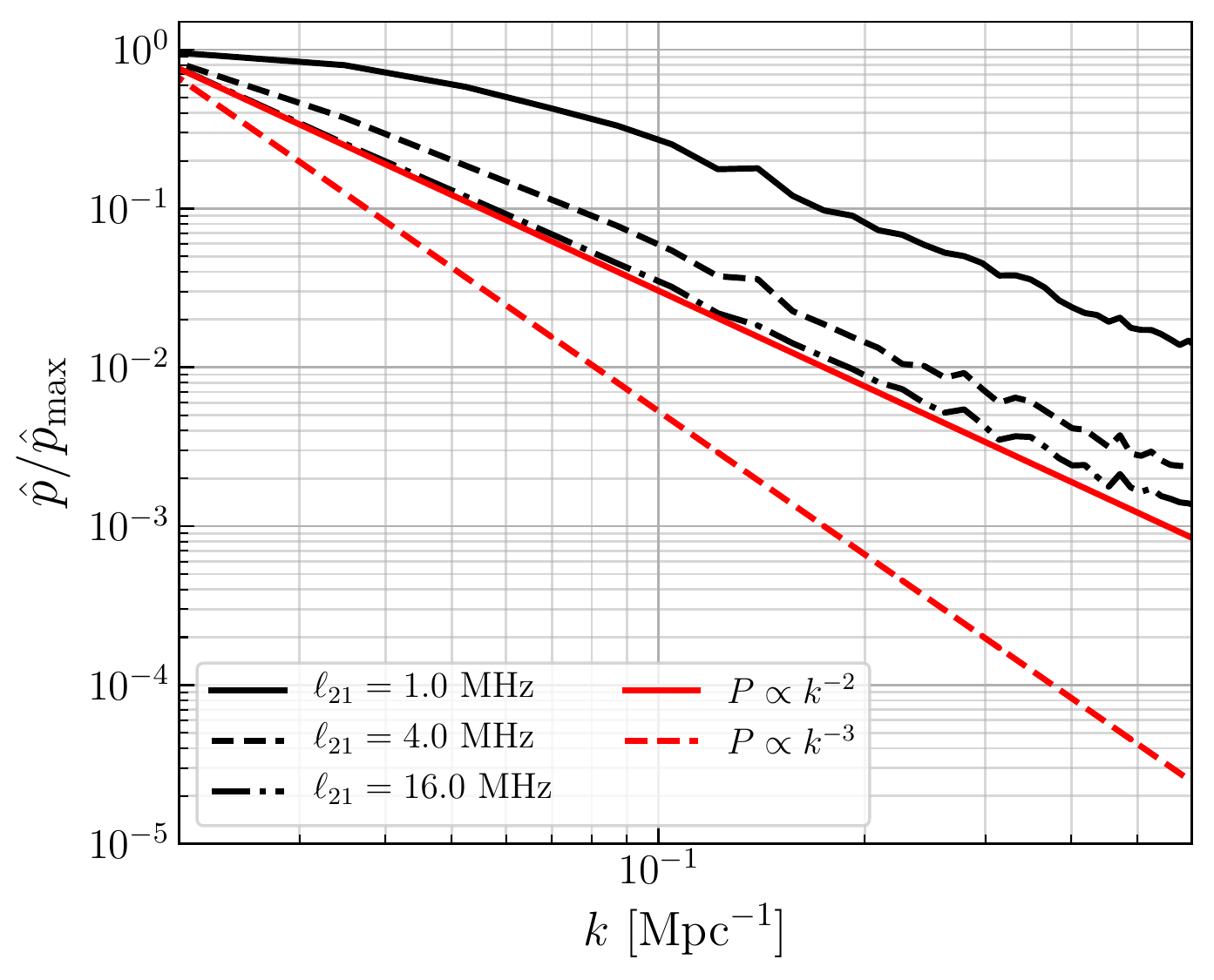}
\caption{Artificially normalized power spectra of an exponential EoR covariance model with different $\ell_{21}$ parameters (black), compared to a power law model (red). Many of the extreme models ruled out by 21\,cm upper limits follow a $\Delta^2\propto k$ or $\Delta^2\propto k^0$ dependence \citep{Ghara2020}, which is equivalent to a $P\propto k^{-2}$ and $P\propto k^{-3}$ dependence, respectively. While an exponential EoR covariance comes close to $P\propto k^{-2}$ when taking $\ell_{21}\rightarrow\infty$, it simply cannot reproduce a $P\propto k^{-3}$ model, suggesting a need to make the EoR model more flexible.}
\label{fig:eor_covariance_power_law}
\end{figure}

The results of the previous section suggests that the power spectrum upper limits of \citetalias{Mertens2020} may be underreporting the EoR power spectrum for $k < 0.08\ {\rm Mpc}^{-1}$.
This is largely due to the fact that, as described in \autoref{sec:bias_correction_math}, their power spectrum normalization is fairly sensitive to any mismatch between the true and adopted EoR covariance.
Because their EoR covariance model is restricted in its degrees of freedom (two hyperparameters of an exponential function), it can only represent certain kinds EoR models.
Although \citetalias{Mertens2018} demonstrate that an exponential function can represent some of the fiducial models from an EoR simulation, this is not true for many other unconstrained models.

Model selection or parameter inference is generally performed with the dimensionless power spectra,
\begin{align}
\label{eq:dsq}
\Delta^2(k) = P(k)\frac{k^3}{2\pi^2},
\end{align}
which has units of mK$^2$.
Thermal noise, being nearly flat across $k$ in $P(k)$, acquires a $k^3$ dependence in $\Delta^2$.
In many fiducial models, the $\Delta^2_{21}$ has either a $\Delta^2\propto k^1$ or a $\Delta^2\propto k^0$ dependence \citep[e.g.][]{Greig2015a}.
This means that the lowest $k$ modes are generally the ones with the largest signal-to-noise, and therefore dominate the data likelihood.
Any amount of signal loss on these modes will dramatically alter the likelihood function.

Furthermore, these kinds of $\Delta^2$ scalings translate to a scaling of $P(k)\propto k^{-2}$ and $P(k)\propto k^{-3}$, respectively.
In \autoref{fig:eor_covariance_power_law}, we show the scaling in $P(k)$ for the exponential EoR covariance adopted in the previous sections (artificially normalized), with an increasing length scale parameter (black).
We also plot a power law response of $P(k)\propto k^{-2}$ and $P(k)\propto k^{-3}$ (red), showing that when we take $\ell_{21}\rightarrow\infty$, the exponential covariance can reproduce a $P(k)\propto k^{-2}$ fairly well.
However, the covariance is unable to adequately describe a $P(k)\propto k^{-3}$ EoR model, regardless of the length scale parameter.
This raises the concern that the lowest $k$ modes of the \citetalias{Mertens2020} limits at $k\sim0.05\ {\rm Mpc}^{-1}$ and $k\sim0.07\ {\rm Mpc}^{-1}$ may be under-reported when compared to the flat and nearly flat $\Delta^2(k)$ models ruled out by \citet{Ghara2020, Mondal2020, Greig2020}.
We therefore advocate for the quadratic estimator-based GPR foreground subtraction technique (without bias subtraction), which is more robust to the effect of mismatched covariances.

\section{Conclusions}
\label{sec:conclusions}

In this work we investigated foreground modeling and subtraction for 21\,cm surveys of the EoR using a Gaussian process (GP) data model.
Previous works have argued that GP foreground subtraction (GPR-FS) may enable a clean separation of foregrounds from the EoR, particularly at the low Fourier $k$ modes where the EoR signal-to-noise is the strongest \citep{Mertens2018}.
A number of 21\,cm EoR analyses have since utilized this technique in their pipelines \citep{Gehlot2018, Ghosh2020, Mertens2020}.
Our goal was to re-visit this topic and to cast GPR-FS in terms of the quadratic estimator (QE) framework in order to put it on stronger theoretical footing.
Incidentally, we showed that GPR-FS is in fact closely related to the widely studied optimal quadratic estimator formalism.

We first showed that, in the limit that we know the true covariances of our data, GPR-FS can help to suppress foreground contamination and recover some of the low $k$ modes (barring additional systematics).
However, we also showed that doing so distorts the window functions of the low $k$ band powers, which require decorrelation if one is to set robust limits or claim an EoR detection at those modes.
Next, we introduced extensions to GPR-FS that improve its ability to suppress foregrounds and allow it to handle the real-world issue of missing or nulled data due to radio frequency interference.

Lastly, we looked at the implementation of GPR-FS in recent upper limits set by LOFAR \citep{Mertens2020}.
We showed that their normalization technique is susceptible to EoR signal loss when the EoR covariance is mismatched from the adopted covariance model, demonstrated with simulated power spectrum recovery tests where we relax the assumption of a known covariance and attempt to regress for it directly from the data.
While they take measures to validate their results by varying their data covariance models and associated hyperparameters, this did not account for many of the kinds of EoR models subsequently ruled out by astrophysical constraints \citep{Ghara2020, Mondal2020, Greig2020}.
Because we show that the \citetalias{Mertens2020} estimator is sensitive to EoR signal loss in the event of a mismatched EoR covariance, we argue that many of these models may have been falsely ruled out.
We note that the quadratic estimator also suffers from EoR signal loss when applying its bias subtraction in the event of a mismatched foreground covariance, but when this is excluded the QE is able to report robust (albeit less sensitive) upper limits across all $k$ modes probed even with a mismatched covariance.
More to the point, the standard QE formalism requires us to know the EoR covariance for the estimator to be optimal, but we do not need to know the EoR covariance for the estimator to be unbiased.
Future work will apply GPR-FS to experiment-specific data simulations with realistic foreground and systematic contamination to explore its ability to suppress these components in a more realistic context.

\section*{Acknowledgements}

The authors would like to thank Florent Mertens, Gianni Bernardi, Abhik Gosh, Miguel Morales, and Aaron Ewall-Wice for helpful discussions related to this work.

NK acknowledges support from the MIT Pappalardo Fellowship.
AL acknowledges support from the New Frontiers in Research Fund Exploration grant program, a Natural Sciences and Engineering Research Council of Canada (NSERC) Discovery Grant and a Discovery Launch Supplement, the Sloan Research Fellowship, the William Dawson Scholarship at McGill, as well as the Canadian Institute for Advanced Research (CIFAR) Azrieli Global Scholars program.

\subsubsection*{Software}
This work relied heavily on publicly available and open-sourced community Python software, including \texttt{numpy} \citep{2020NumPy-Array}, \texttt{scipy} \citep{scipy2020}, \texttt{scikit-learn} \citep{Pedregosa2012}, and \texttt{emcee} \citep{Foreman-Mackey2013}.

\subsubsection*{Data Availability}
Data used in this work may be made available upon reasonable request to the corresponding author.



\bibliographystyle{mnras}
\bibliography{eor_references} 



\appendix

\section{Power Spectrum Normalization and Bias}
\label{sec:bias_correction_math}

In this section we discuss the differences between power spectrum normalization and bias, looking at how this is defined in the standard QE estimator and how it is defined in \citetalias{Mertens2020}'s estimator.
In summary, we show that because \citetalias{Mertens2020} choose to deal with power spectrum normalization via a bias correction procedure, their estimator is particularly sensitive to inadvertent signal loss when the EoR covariance is misestimted.
This is in part driven by the assumption they make in their normalization step that the adopted covariance model is equal to the true covariance of the data.

First, we define the relationship between the unnormalized band power estimate, $\hat{\q}$, and the normalized band power estimate, $\hat{\p}$, within the quadratic estimator framework.
The unnormalized band powers are given as
\begin{align}
\label{eq:unnorm_q}
\hat{q}_\alpha = \frac{1}{2}\d^\dagger\R^\dagger\C_{,\alpha}\R\d,
\end{align}
where the $\R$ and $\C_{,\alpha}$ matrices are defined in the same wasy as \autoref{sec:qe}.
Taking its expectation value and utilizing the cyclic property of the trace, we arrive at
\begin{align}
\label{eq:unnorm_q_avg}
\langle\hat{q}_\alpha\rangle &= \frac{1}{2}\tr[\R\langle \d\d^\dagger\rangle\R^\dagger\C_{,\alpha}] \nonumber \\ &= \frac{1}{2}\tr[\R\C\R^\dagger\C_{,\alpha}] \nonumber \\
&= \frac{1}{2}\tr[\R(\Cn+\Cfg)\R^\dagger\C_{,\alpha}] + \sum_\beta\frac{1}{2}\tr[\R\C_{,\beta}\R^\dagger\C_{,\alpha}]p_\beta \nonumber \\
&= \frac{1}{2}\tr[\R(\Cn+\Cfg)\R^\dagger\C_{,\alpha}] + \sum_\beta H_{\alpha\beta}p_\beta,
\end{align}
where in the last line we recognized the coefficient of $p_\beta$ as $H_{\alpha\beta}$.
Note that going from the first to the second line we substituted in $\langle \d\d^\dagger\rangle=\C$, where $\C$ is the true (but not necessarily known) covariance of the data.
Following \autoref{sec:qe}, an estimate of the normalized band power can therefore be written as
\begin{align}
\label{eq:norm_bp}
\hat{p}_\alpha = \sum_\beta M_{\alpha\beta}\hat{q}_\beta - \hat{b}_\alpha.
\end{align}
Recall that our definition of a normalized estimator is one that satisfies
\begin{align}
\label{eq:norm_condition}
\langle\hat{\p}\rangle = \W\p,
\end{align}
where the rows of $\W$ must sum to unity.
For completeness, we can show that the expectation value of \autoref{eq:norm_bp} is indeed properly normalized
\begin{align}
\label{eq:avg_phat}
\langle \hat{p}_\alpha \rangle &= \sum_\beta M_{\alpha\beta}\langle\hat{q}_\beta\rangle - \langle\hat{b}_\alpha\rangle \nonumber \\
&= \sum_{\beta\gamma} M_{\alpha\beta}H_{\beta\gamma}p_\gamma + \tr[(\Cn + \Cfg)\E_\alpha] - \langle\hat{b}_\alpha\rangle \nonumber \\
&= \sum_\gamma W_{\alpha\gamma}p_\gamma + \tr[(\Cn + \Cfg)\E_\alpha] - \langle\hat{b}_\alpha\rangle.
\end{align}
From the last line, we see that for the 21\,cm component of our estimated band powers (i.e. $p_\gamma$), we simply require an $\M$ matrix to satisfy $\sum_{\beta\gamma} M_{\alpha\beta}H_{\beta\gamma} = 1$.
However, a separate condition ensures that the power spectra are free of excess contaminants, which is the condition that our estimated bias term be itself unbiased, or
\begin{align}
\label{eq:bias_expectation}
\langle\hat{b}_\alpha\rangle = \tr[(\Cn + \Cfg)\E_\alpha].
\end{align}
Note that in the form that we have derived it, $\langle \hat{b}_\alpha\rangle$ is purely positive.
This strict condition indeed requires us to know the noise and foreground covariance, which in practice is never achieved exactly.
While one can eliminate the noise bias from a practical estimator \citep{Dillon2014}, the foreground bias term is particulary tricky, and can actually lead to signal loss if it is misestimated \citep{Dillon2014, Cheng2018}.
A conservative choice is to set $\hat{b}_\alpha=0$, which will leave in some of the foreground contamination, but will never underpredict the EoR signal because we have constructed the foreground bias as a purely positive excess.

\begin{figure*}
\centering
\includegraphics[width=\linewidth]{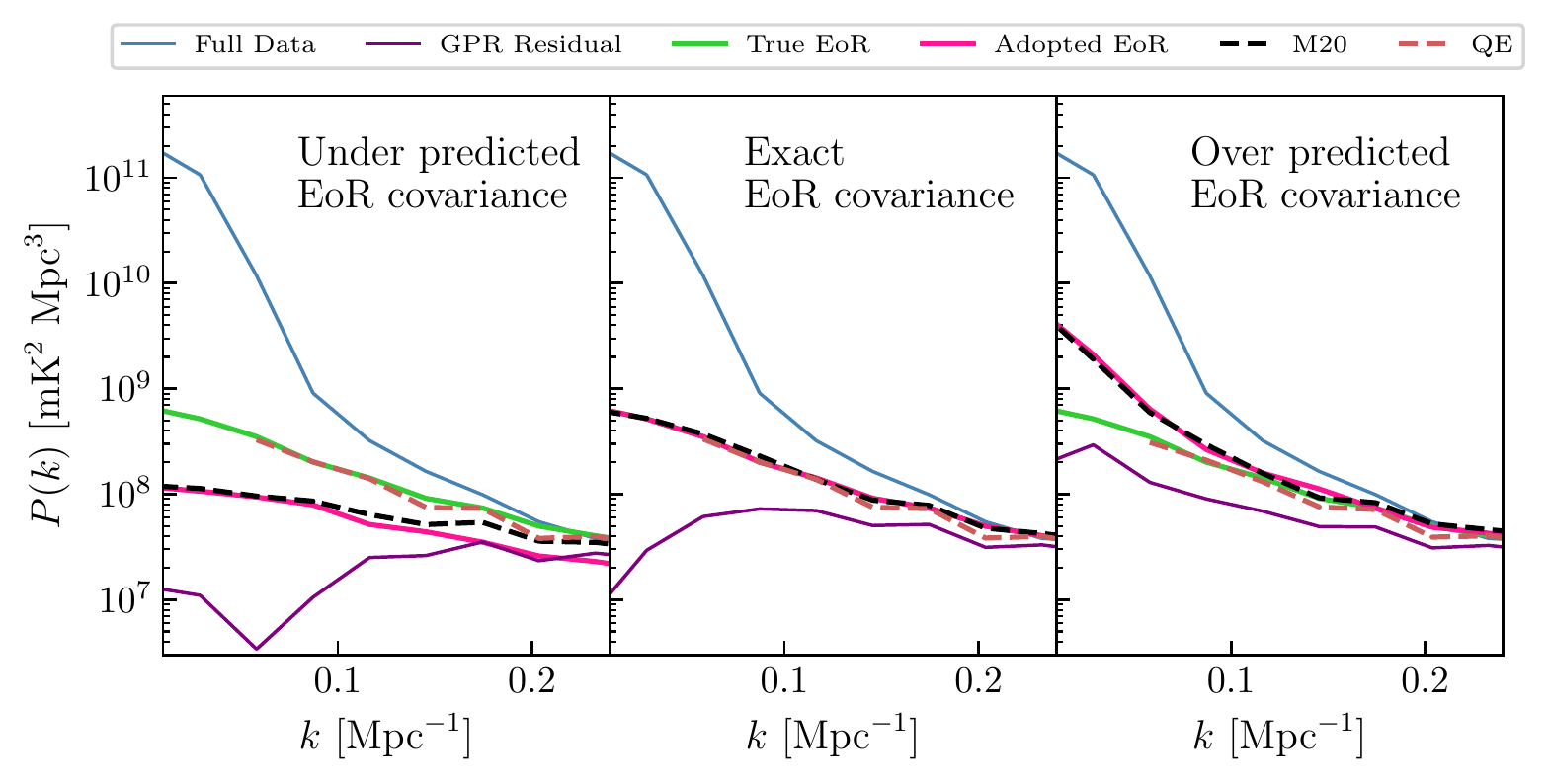}
\caption{LOFAR power spectrum normalization test (without the excess noise component) assuming we know the exact data covariance for the foregrounds and noise, but have misestimated the EoR covariance. The power spectrum of the \emph{adopted} EoR covariance is plotted in pink, which in the left and right panels do not match the power spectrum of the \emph{true} EoR covariance (green). We can see that at low $k$ modes the normalized LOFAR power spectra (black dashed) closely track the EoR model given by the adopted covariance, not the true EoR model. This trend is not seen for the QE implementation of GPR-FS (red dashed), which is constructed with an $\M=\H^{-1/2}$ normalization, has the foreground bias term subtracted off (because we know it exactly), and is truncated below $k<0.03\ {\rm Mpc}^{-1}$ because its window functions are poorly localized.}
\label{fig:lofar_pspec_adopted_true}
\end{figure*}

A different approach is taken by \citetalias{Mertens2020}.
To summarize their Sec. 3.2.2, the covariance of the residual vector after GPR foreground removal can be shown to be
\begin{align}
\label{eq:res_covariance}
\langle\r\r^\dagger\rangle &= \R_{\operatorname{GPR-FS}}\langle\d\d^\dagger\rangle\R_{\operatorname{GPR-FS}}^\dagger \nonumber \\
&= [\Cn + \Cto] - \Cfg + \Cfg[\Cfg + \Cn + \Cto]^{-1}\Cfg \nonumber \\
&= [\Cn + \Cto] - \Cov[\f],
\end{align}
where they make the key assumption that their covariance models are equal to the true covariance of the data in order to simplify the expression.
\autoref{eq:res_covariance} then tells us that in order to unbias the covariance of our measured residual, we should sum it with the covariance of the GPR foreground estimate, $\Cov[\f]$.
In practice, \citetalias{Mertens2020} implement this correction for their power spectrum estimator using a Monte Carlo procedure:
they draw many random realizations of the data from the foreground covariance (\autoref{eq:fg_cov_conditional}) and average their power spectra before directly summing the result with the power spectrum of the residual vector.
We can see already that the idea of normalizing the estimator by adding a bias term runs counter to the QE normalization scheme in \autoref{eq:avg_phat}, where the normalization is multiplicative, and the bias is a purely positive quantity that is subtracted off.

In adopting $\R = \R_{\rm GPR-FS}$, we see that \autoref{eq:unnorm_q} is simply the (unnormalized) power spectrum of the GPR-FS residual data vector: $\hat{q}_\alpha = \tfrac{1}{2}\r^\dagger\C_{,\alpha}\r$.
Taking the expectation value of the unnormalized band power and substituting in \autoref{eq:res_covariance}, we get
\begin{align}
\langle\hat{q}_\alpha\rangle &= \frac{1}{2}\langle \r^\dagger\C_{,\alpha}\r\rangle \nonumber \\ 
&= \frac{1}{2}\tr[\langle\r\r^\dagger\rangle\C_{,\alpha}] \nonumber \\
&= \frac{1}{2}\tr[\Cn\C_{,\alpha}] - \frac{1}{2}\tr[\Cov[\f]\C_{,\alpha}] + \sum_\beta\frac{1}{2}\tr[\C_{,\alpha}\C_{,\beta}]p_\beta.
\end{align}
Multiplying by the $\M$ matrix and casting it into the form of \autoref{eq:avg_phat}, we get
\begin{align}
\label{eq:M20_bias_corr}
\langle\hat{p}_\alpha\rangle &= \sum_\beta M_{\alpha\beta}\langle\hat{q}_\beta\rangle  - \langle\hat{b}_\alpha\rangle  \nonumber \\
&= \sum_{\beta\gamma}\frac{1}{2}M_{\alpha\beta}\tr[\C_{,\beta}\C_{,\gamma}]p_\gamma \nonumber \\
&\hspace{.4cm} + \sum_\beta \frac{1}{2}M_{\alpha\beta}\left(\tr[\Cn\C_{,\beta}] - \tr[\Cov[\f]\C_{,\beta}]\right) - \langle\hat{b}_\alpha\rangle.
\end{align}
The result is a different tradeoff between normalization and bias.
Here, the $\H$ matrix is more compact: for a simple $\C_{,\beta}$ matrix containing only Fourier modes, then $H_{\beta\gamma}$ can be shown to be diagonal.
However, we can also see that the bias term is now not necessarily positive.
In fact, for an estimator with no noise bias, we can see that $\langle\hat{b}_\alpha\rangle$ will in fact be purely negative.
Furthermore, because $\Cov[\f]$ is dependent on the full data covariance, the bias term is now itself dependent on the adopted 21\,cm covariance.
We argue that this is a precarious tradeoff, as we noted before that bias correction in the QE has the potential for inducing signal loss.

To demonstrate the difference these conventions have on the resultant GPR-FS power spectra, we run a simple test on the simulation products of \autoref{sec:lofar} (excluding the excess noise component to improve our SNR on the EoR signal).
In our test, we assume we know the exact covariance of the foregrounds and noise components of the data, but vary the EoR covariance slightly such that it overpredicts or underpredicts the EoR signal at low wavenumbers.
This is shown in \autoref{fig:lofar_pspec_adopted_true}, where the left panels shows the case of under prediction, the right panel shows the case of over prediction, and the middle panel is the case where the adopted and true EoR covariance are the same.
The takeaway is that the \citetalias{Mertens2020} normalization (black dashed) closely follows the power spectrum from the adopted EoR covariance (pink) at low $k$, illuminating how susceptible the normalization scheme is specifically to a misestimated EoR covariance.
On the other hand, the QE normalization (red dashed), which includes a bias subtraction because we have assumed we know the true foreground covariance, does not demonstrate this sensitivity, indicative of the fundamentally different properties of the two normalization schemes.


\bsp	
\label{lastpage}
\end{document}